\documentclass[aps,prfluids,tightenlines,preprint]{revtex4-2}

\usepackage{graphicx}
\usepackage[dvipsnames]{xcolor}
\usepackage{amssymb}
\usepackage{amsmath}
\usepackage[dvipsnames]{xcolor}
\usepackage{subcaption}
\usepackage{array}
\newcommand{\lr}[1]{\left(#1\right)}
\newcolumntype{L}{>{\centering\arraybackslash}m{1.5cm}}

\begin{document}

\title{Coherent Solutions and Transition to Turbulence in Two-Dimensional Rayleigh-B\'enard Convection}

\author{Parvathi Kooloth}
\affiliation{Department of Mathematics, University of Wisconsin-Madison, Madison, WI 53706, USA}
\author{David Sondak}
\email{dsondak@seas.harvard.edu}
\affiliation{Institute for Applied Computational Science, Harvard University, Cambridge, MA 02138, USA}
\author{Leslie M. Smith}
\affiliation{Department of Mathematics and Department of Engineering Physics, University of Wisconsin-Madison, Madison, WI 53706, USA}

\begin{abstract}
For two-dimensional (2D) Rayleigh-B\'{e}nard convection, classes of unstable, steady solutions were previously computed using numerical continuation \cite{waleffe2015heat,sondak2015optimal}. The `primary' steady solution bifurcates from the conduction state at 
$Ra \approx 1708$, and has a characteristic aspect ratio (length/height) of
approximately $2$.
The primary solution corresponds to one pair of counterclockwise-clockwise convection rolls with a temperature updraft in between and an adjacent downdraft on the sides. By adjusting the horizontal length of the domain, \cite{waleffe2015heat,sondak2015optimal} also found steady, maximal heat transport solutions, with characteristic aspect ratio
less than $2$ and decreasing with increasing $Ra$. Compared to the primary solutions, optimal heat transport solutions have modifications to boundary layer thickness, the horizontal length scale of the plume, and the structure of the downdrafts. The current study establishes a direct link between these (unstable) steady solutions and transition to turbulence for $Pr = 7$ and $Pr = 100$. For transitional values of $Ra$, the primary and optimal-heat-transport solutions both appear prominently in appropriately-sized sub-fields of the time-evolving temperature fields.  
For $Ra$ beyond transitional, our data analysis 
shows persistence of the primary solution for $Pr = 7$, while the optimal heat transport solutions are more easily detectable for $Pr = 100$. In both cases $Pr = 7$ and $Pr = 100$, the relative prevalence of primary and optimal solutions is consistent with the $Nu$ vs.\ $Ra$ scalings for the numerical data and the steady solutions.
%scaling of the Nusselt number with $Ra$ is in close agreement with the scaling of the primary solution.
\end{abstract}

\maketitle 

\section{Introduction}
Thermal convection is heat transfer in liquids and gases resulting from buoyancy-driven fluid motion, and is fundamental for establishing circulations in the earth's atmosphere and oceans, planetary mantle dynamics, stellar evolution, as well as a broad range of engineering applications \cite[]{lappa2009thermal}.  Rayleigh-B\'enard convection (RBC) refers to convection generated by simplified dynamical equations and idealized boundary conditions.  The simplified equations are called the Oberbeck-Boussinesq approximation, describing viscous fluids that are mechanically incompressible but thermally expansible \cite[see, e.g.,][and references therein]{rajagopal1996oberbeck}.
The domain is confined between two parallel plates, heated from below and cooled from above such that the bottom plate is always warmer than the top plate by a fixed temperature difference.  The Rayleigh-B\'enard setup presents a framework that is at the same time experimentally realizable, analytically tractable and numerically accessible \cite[see][for  overviews]{ahlers2009heat,chilla2012new}.  Its prominence in the literature stems partly from its practical importance, and partly from the fact that Rayleigh-B\'enard convection represents a rich nonlinear dynamics to test physical reasoning and mathematical techniques. 

In the Rayleigh-B\'{e}nard convection problem, the Rayleigh number $Ra$ and the Prandtl number $Pr$ are nondimensional parameters that control the flow dynamics: $Ra$ characterizes the relative strength of buoyancy-driven inertial forces to viscous forces and $Pr$ is the ratio of the kinematic viscosity to the thermal diffusivity.  Furthermore, a key nondimensional diagnostic parameter is the Nusselt number $Nu$, measuring the vertical heat transport for a given temperature difference between the plates. Over the last several decades, there has been an intense focus on the scaling behavior of $Nu$ on $Ra$ and $Pr$. This relationship has been explored via experiments \cite{niemela2000turbulent, chavanne2001turbulent, niemela2006turbulent,  ahlers2009heat, he2012heat, he2012transition, bouillaut2019transition}, scaling laws \cite{malkus1954heat, priestley1954convection, kraichnan1962turbulent, spiegel1963generalization, howard1963heat, busse1969howard, grossmann2000scaling, grossmann2002prandtl}, 
rigorous upper bounds on the allowable heat transport~\cite{doering1994variational, constantin1995variational, doering1996variational, kerswell2001new, ding2019exhausting}, and numerical simulations~\cite{bailon2010aspect,emran2015large,waleffe2015heat,sondak2015optimal,motoki2018maximal,fonda2019deep,iyer2020classical, kawano2020ultimate, wang2020multiple, sondak2020high}. The two competing scaling behaviors that emerge from this body of work are $Nu\propto Ra^{1/3}$ and $Nu\propto Ra^{1/2}$. The $1/3$ scaling, often referred to as the classical scaling, emerges from physical arguments~\cite{malkus1954heat, priestley1954convection} involving independence of top and bottom boundary layers~\cite{priestley1954convection} as well as marginal stability of the boundary layers~\cite{malkus1954heat, howard1963heat}. The $1/2$ scaling, recently referred to as the ultimate regime, emanates from mixing length theories of turbulence~\cite{spiegel1963generalization, kraichnan1962turbulent}. There has been a flurry of recent activity on the possible transition from scaling $Nu \propto Ra^{1/3}$ to scaling $Nu \propto Ra^{1/2}$, {e.g.}\ experiments by \cite{niemela2000turbulent, xia2002heat, he2012transition, bouillaut2019transition}.  Several studies have also been aimed at accessing the regime for $Ra > 10^{13}$, and at verifying (or not) the existence of the scaling $Nu \propto Ra^{1/2}$~\cite{ahlers2017ultimate, bouillaut2019transition, doering2019thermal, stevens2018direct, zhu2018transition, doering2019absence, doering2020turning, iyer2020classical}.  Such considerations are beyond the scope of our work in this paper. We note that the flow Reynolds number can be expressed in terms of $Ra$ and $Pr$, and that the precise relationship is also of interest to researchers~\cite{grossmann2000scaling,grossmann2004fluctuations,stevens2013unifying}. Interestingly, rigorous upper bounds on heat transport are sensitive to the boundary conditions, being bounded by $1/2$ in the no-slip case~\cite{doering1996variational} and by $5/12$ in the 2D free-slip case~\cite{whitehead2011ultimate,hassanzadeh2014wall}. We also mention that substantial effort has been devoted to determining the effect of $Pr$ on the heat transport~\cite{whitehead2014rigorous, choffrut2016upper, otto2017rigorous,doering2006bounds,wang2007asymptotic,otto2011improved}. Finally, the Grossmann-Lohse (GL) framework~\cite{grossmann2000scaling,grossmann2002prandtl} introduces nonlinear relationships for $Nu\lr{Ra,Pr}$ and $Re\lr{Ra,Pr}$ with six free parameters that have been fit to experimental data~\cite{stevens2013unifying}.

In \cite{waleffe2015heat,sondak2015optimal}, the 2D Oberbeck-Boussinesq equations were solved numerically to find steady solutions.  All of the computed solutions consist of a symmetric pair of hot and cold plumes emanating from the bottom and top boundary layers, respectively, and extending nearly wall-to-wall.  The `primary' solution bifurcates from the conduction state at $Ra \approx 1708$, and has a characteristic aspect ratio of %$l/h \approx 2$, where $l$ ($h$) is the half-length (half-height) in the horizontal (wall-normal) direction. 
$\Gamma_{prim} = L_{prim}/{H_{prim}}\approx 2$, where $L_{prim}$ and $H_{prim}$ are the length and height of the domain, respectively. 
By adjusting the horizontal length of the domain, $L$,~\cite{waleffe2015heat,sondak2015optimal} also looked for maximal heat transport solutions, with characteristic aspect ratio $\Gamma_{max} < 2$, decreasing with increasing $Ra$. For each value of $Ra$ and $Pr$ considered, they computed solutions that maximized heat transport, referred to as the `optimal solution' (with highest $Nu$). For higher $Ra$, two local maxima emerged with different aspect ratios, and $Pr$ determined which maximum was the global optimal heat transport solution~\cite{sondak2015optimal}. Best fit scaling relations 
$Nu -1 \propto Ra^\beta$ produce 
$\beta \approx 0.315$ ($\beta \approx 0.311$) for the optimal solution with $Pr = 7$ ($Pr = 100$), and $\beta \approx 0.28$ ($\beta \approx 0.227$) for the primary solution with $Pr=7$ ($Pr=100$).
%see Figure~\ref{fig:Nu-Ra-AR10} (fits start at $Ra \approx 10^6$).

\begin{figure}[h!]
  \centering
  \includegraphics[width=\textwidth]{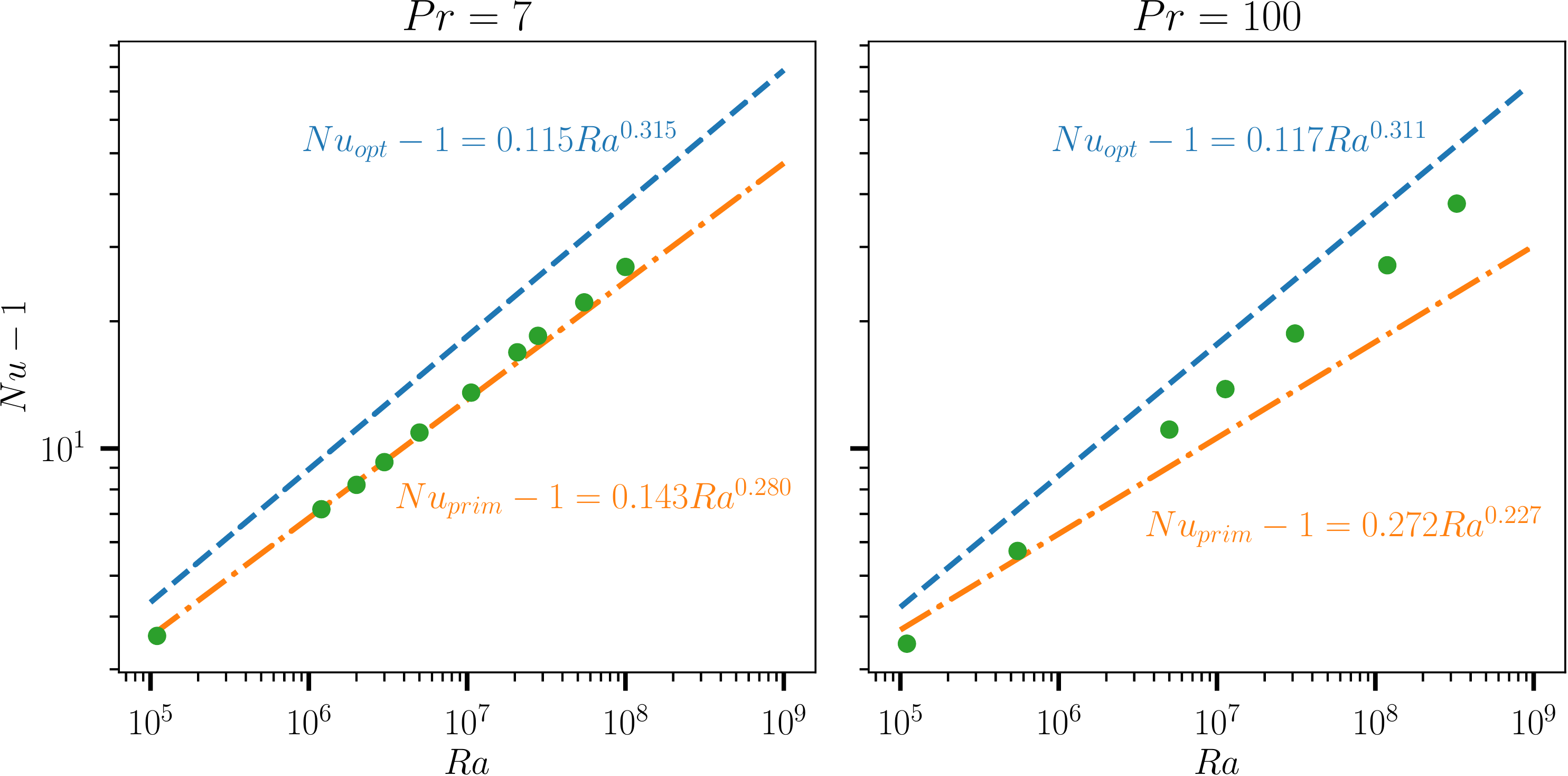}
  \caption{$Nu$ vs.\ $Ra$ scaling for 2D Rayleigh-B\'{e}nard convection: $Pr = 7$ (left), $Pr = 100$ (right); symbols are simulation data in a domain of aspect ratio 
  %$\Gamma =L/H = 10$; 
  $\Gamma=10$; dashed lines are best-fit scaling relations for the optimal solutions; dash-dotted lines are the best-fit scaling relations for the primary solutions. Scaling for simulation data is $Nu = 0.105Ra^{0.301}$ for $Pr=7$ and $Nu = 0.119Ra^{0.293}$ for $Pr=100$.  
  %All fits start at $Ra \approx 10^6$ and use the available data for higher $Ra$.  The fit for the primary solutions have been extended beyond the highest available $Ra \approx 5 \times 10^7$ to aid the eye.
  }
  \label{fig:Nu-Ra-AR10}
\end{figure}

The work presented herein aims to establish a direct link between the steady solutions \cite{waleffe2015heat,sondak2015optimal}, and the structures observed in time-evolving Rayleigh-B\'{e}nard flows. As motivation for the study, Figure~\ref{fig:Nu-Ra-AR10} shows the Rayleigh-B\'{e}nard simulation data in a two-dimensional (2D) domain of aspect ratio 
%$\Gamma = L/H = 10$, 
$\Gamma=10$, along with the best-fit scalings for the primary and optimal solutions.
%(fits start at $Ra \approx 10^6$).
With the expection of the optimal fit at $Pr=7$, which starts at $Ra=10^{7}$, all fits start at $Ra \approx 10^6$ and use the available data for higher $Ra$.  The fit for the primary solutions have been extended beyond the highest available $Ra \approx 5 \times 10^7$ to aid the eye.
For both $Pr=7,100$ at low $Ra < 10^7$, the $Nu$ values for the simulation data are closer to the values corresponding to the primary solution.
In the range $10^7 < Ra < 10^9$, the $Nu$ values for the data consistently become higher than for the primary solution, and the optimal solutions provide a tight upper bound on both sets of data.  For $Pr=100$, it is especially evident that $Nu$ values for the data are in between primary and optimal $Nu$ values, with approximate scaling exponent
$0.293$ closer to the optimal exponent $0.311$ than to the primary exponent $0.227$.  

\begin{figure}[h!]
  \centering
  \includegraphics[width=0.85\linewidth]{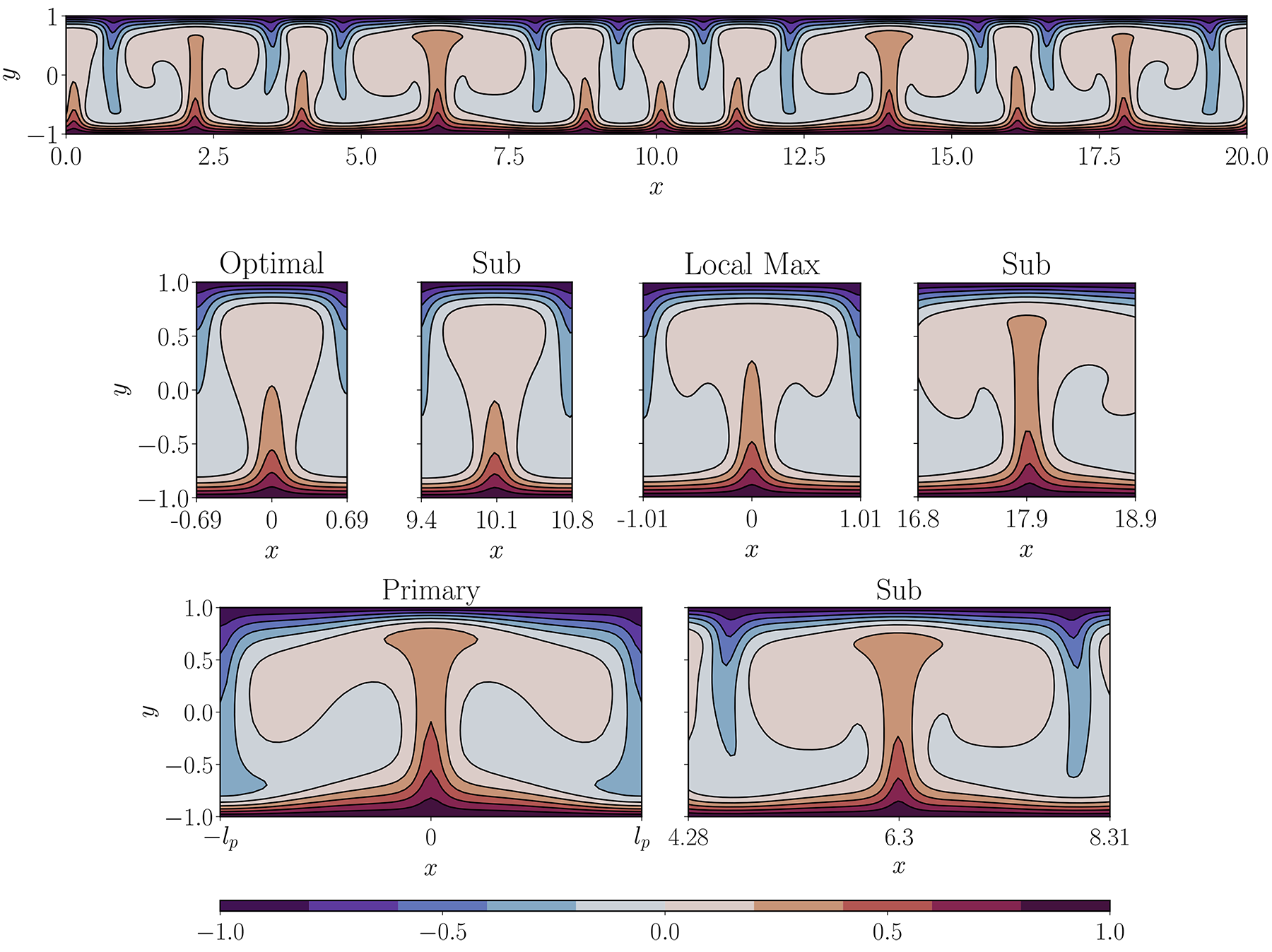}
  \caption{Comparison of temperature fields with $Ra = 1.1 \times 10^5$ and  $Pr = 100$. The top shows a snapshot at time $t=1742$ in a simulation evolving from initial conditions (\ref{eqn:initialconditions}) with random amplitudes (see also Figures~\ref{fig:Nu_t_1E5_100} and \ref{fig:transient-Q_p}).
  The middle left, middle right and bottom panels compare, respectively, the optimal, local maximum and primary solution to different simulation sub-boxes with titles `Sub'.
  The horizontal scale of the primary solution is $L_{prim} = 2l_p$.}
  \label{fig:1E5_100_2}
\end{figure}

Furthermore, Figure~\ref{fig:1E5_100_2} shows the appearance of all three steady solutions --- primary, optimal, and local maximum --- within a single time snapshot of a simulation with domain aspect ratio $\Gamma = 10$, $Ra=1.1\times 10^{5}$, and $Pr=100$. The time $t=1742$ is relatively early in the evolution from initial conditions (\ref{eqn:initialconditions}) with random amplitudes. Notice that these initial conditions do not select a scale in the horizontal direction, and that the horizontal scales of the primary, optimal and local max arise spontaneously from nonlinear interactions.  At this early time $t = 1742$, the flow is quasi-steady, and later transitions to a quasi-periodic, statistically steady state.
In Figure~\ref{fig:1E5_100_2}, the middle left, middle right and bottom panels compare, respectively, the optimal, local maximum and primary solution to different simulation sub-boxes.  In each case, the sub-box has the same horizontal scale as the corresponding steady solution, e.g., the primary solution has horizontal scale $L_{prim}=2l_p\approx 4$.

%Motivated by the observations in Figures~\ref{fig:Nu-Ra-AR10} and~\ref{fig:1E5_100_2}, 

Motivated by Figures~\ref{fig:Nu-Ra-AR10} and~\ref{fig:1E5_100_2}, the purpose of the current investigation is to explore how transition to turbulence in 2D Rayleigh-B\'{e}nard convection is influenced by the primary and optimal steady solutions. We focus on Rayleigh numbers $10^5 < Ra < 10^9$ well above the onset of convection and well below the ultimate regime. In this range, the flow is potentially described as a dynamical systems `repeller' consisting of unstable solutions that are sampled by turbulence trajectories~\cite{gibson2008visualizing}. In such a description, solutions with a small number of unstable directions in phase space would be sampled most frequently.  With increasing $Ra$, one would expect more and more unstable (and unsteady) solutions of increasing complexity in the phase space, thus complicating such an investigation at higher $Ra$. The possibility of describing turbulence as a repeller has been previously explored in the context of wall-bounded shear flows (see~\cite{nagata1990three, waleffe2001exact, gibson2008visualizing, viswanath2008dynamics, hall2010streamwise, eckhardt2007turbulence, chini2016exact} and references therein).

We restrict our investigation to $Pr=7$ and $Pr=100$, in part because their optimal solutions are distinct from one another~\cite{sondak2015optimal}.  The optimal thermal fields of $Pr=7$ cases have coiling arms and 
%thus 
a larger wavelength, as compared to those of $Pr=100$ which are armless and columnar~\cite{waleffe2015heat,sondak2015optimal}. Furthermore, $Pr=7$ has the added benefit of having been investigated extensively in the literature.  Despite the differences in the nature of the optimal solution for $Pr=7$ and $Pr=100$, the scaling behavior of $Nu$ with $Ra$ is almost identical (see Figure~\ref{fig:Nu-Ra-AR10}).  We perform well-resolved numerical simulations and analyze the resulting datasets using two main analysis techniques.  First we consider the spatial correlation between the computed steady-state solutions (primary or optimal) and `windows' of the simulation data with the same aspect ratio as the steady-state solution. Second, we use the singular value decomposition on the windows with high correlation to compare the underlying structures.  Our work fits within a broader class of data-driven approaches to explore structures and dynamics in Rayleigh-B\'{e}nard convection. For example, the dynamic mode decomposition has been used to search turbulence data for unstable periodic orbits~\cite{page2019searching}. Turbulent superstructures with scale separation were identified in~\cite{pandey2018turbulent}. In~\cite{fonda2019deep}, a convolutional neural network was trained and used to analyze turbulent superstructures. Constrained generative adversarial neural networks have been used to learn the underlying distribution of a Rayleigh-B\'{e}nard convection dataset~\cite{wu2020enforcing}.
 
The rest of the paper is organized as follows. Section~\ref{sec:bg} presents the governing equations and the details of the numerical simulations. Section~\ref{sec:method} describes the methodologies we used to detect the footprint of the steady solutions, and the rationales behind them. The results are presented in Section~\ref{sec:results}, followed by  discussion in Section~\ref{sec:conclusions}.

\section{Background}\label{sec:bg}
\subsection{Governing Equations}
In Rayleigh-B\'{e}nard convection, a fluid under the influence of gravity is contained between two horizontal plates separated by a distance $H$ such that the lower plate is held at a higher temperature ($T_{w}$) than the upper plate ($-T_{w}$). Buoyancy effects are incorporated into the Navier-Stokes equations via the Oberbeck-Boussinesq approximation~\cite{oberbeck1879warmeleitung,boussinesq1903theorie,chilla2012new, chandrasekhar2013hydrodynamic}, wherein density is assumed to vary linearly with temperature in the buoyancy term. In particular, the density variations in the buoyancy term have the form $\rho\lr{T} = \rho_{0} - \rho_{0}\alpha_{V}\lr{T - T_{0}}$ where $\rho_{0}=\rho\lr{T_{0}}$, $T_{0}$ is a reference temperature, and $\alpha_{V}$ is the coefficient of volume expansion of the fluid. In the conduction state, the fluid is quiescent and the temperature conduction profile varies linearly with $y$~\cite{drazin2004hydrodynamic}. For plates situated at $y=\pm h$ such that $H=2h$, the conduction profile is $T_{c} = -\Delta T y / H$ where $\Delta T$ is the temperature difference between the bottom and top plate. In the present work, we restrict our investigation to two-dimensional flows. We work with a nondimensional form of the equations where temperature is scaled by $\Delta T/2$, length by $h$, time by the free fall time $t_{f}=\sqrt{h / \lr{g\alpha_{V}\Delta T/2}}$, velocity by the free fall velocity $U_{f} = h / t_{f}$, and pressure by a dynamic pressure $\rho_{0}U_{f}^{2}$. All variables, unless explicitly stated, are understood to be nondimensionalized. The nondimensional governing equations are, 
\begin{align}
   &\frac{\partial \mathbf u}{\partial t} +  \mathbf{u}\cdot\nabla \mathbf{u} = -\nabla P + \nu_{\star}\nabla^2 \mathbf{u} + \Theta \mathbf{\widehat{y}} \label{eq:mom} \\
   &\nabla\cdot\mathbf{u} = 0 \label{eq:cont} \\
   &\frac{\partial \Theta}{\partial t} + \mathbf{u}\cdot\nabla \Theta - v = \kappa_{\star}\nabla^2 \Theta, \label{eq:energy}
\end{align}
where $\mathbf{u} = \left(u,v\right)$ is the velocity, $\Theta = T - T_{c}$ is the temperature departure from the conduction state, and $\widehat{\mathbf{y}}$ is the unit vector $(0,1)$ in the direction perpendicular to the bottom wall. The nondimensional parameters $\nu_{\star}$ and $\kappa_{\star}$ are related to the classical $Ra$ and $Pr$ by,
\begin{align}
  \nu_{\star} = \lr{\frac{16 Pr}{Ra}}^{1/2}, \quad \kappa_{\star} = \lr{\frac{16}{Ra Pr}}^{1/2}
\end{align}
where,
\begin{align}
  Ra = \frac{g\alpha_{V}\Delta T H^{3}}{\nu\kappa}, \quad Pr = \frac{\nu}{\kappa}.
\end{align}
The additional dimensional quantities appearing in the definitions of $Ra$ and $Pr$ are the acceleration due to gravity $g$, the kinematic viscosity $\nu$, and the thermal diffusivity $\kappa$. We note that $P$ in~\eqref{eq:mom} is a modified pressure equal to $P_{m} + \lr{2/\lr{\alpha_{V}\Delta T} + T_{0}}y + y^{2}/2$ where $P_{m}$ is the nondimensional mechanical pressure appearing in the momentum equation and $T_{0}$ is a nondimensional reference temperature. As a final note, in this nondimensionalization the conduction profile is $T_{c} = -y$.

No-slip and fixed temperature boundary conditions are imposed on the plates such that
\begin{align}
  \mathbf u \left(x,\pm 1,t\right) = 0,\quad
  \Theta \left(x,\pm 1,t\right) = 0. \label{bc:topbottom}
\end{align}
The domain extends from $x=0$ to $x=L$ in the horizontal direction, with periodic boundary conditions,
\begin{align}
  \mathbf{u}\left(0, y, t\right) = \mathbf{u}\left(L, y, t\right),\quad
  \Theta\left(0,y,t\right) = \Theta\left(L,y,t\right).
\end{align}
Here $L$ represents the domain length nondimensionalized by $h$ so that the dimensional domain has length $hL$.

\begin{figure}[h!]
  \centering
  \includegraphics[width=\textwidth]{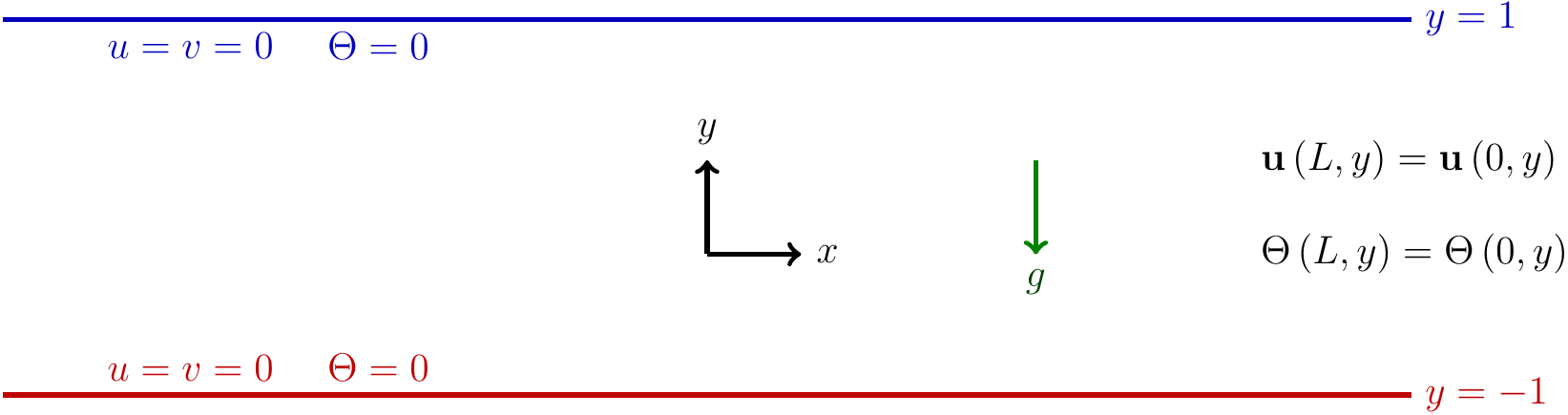}
  \caption{Set-up for Rayleigh-B\'{e}nard convection}
  \label{fig:set_up}
\end{figure}

Figure~\ref{fig:set_up} depicts the problem configuration.
In this geometry, and for the selected nondimensionalization, the domain aspect ratio is $\Gamma = L/2$. Simulations in this domain are compared to the optimal and primary solutions, which extend from $-l_{\text{opt}}$ to $l_{\text{opt}}$ and $-l_{prim}$ to $l_{prim}$, respectively. The horizontal domain for the optimal and primary structures is therefore $L=2l$ where $l$ is the half-width of the structure being considered. In subsequent sections, the half-width of a turbulent plume will be denoted by $l$ and the context will make it clear if the plume is an optimal, primary, local max or turbulent structure. 
%We further note that the aspect ratio can be written $L/H = l/h$.

Before the onset of convection, the heat transfer depends linearly on the temperature difference between the plates. This linear relation breaks down at the onset of thermal instability as the convective modes of heat transport activate~\cite{schmidt1935instability}. The conduction solution is found to be linearly unstable when $Ra$ exceeds a critical value of $\approx 1707.8$ for no-slip boundaries~\cite{drazin2004hydrodynamic,chandrasekhar2013hydrodynamic}. After convection sets in,
the dimensional flux through each of the top and bottom boundaries is given by
\begin{align}
  \mathcal{Q}_{\pm}(t) = -\kappa \left.\frac{d\overline{T}}{dy}\right|_{y=\pm h}
  \label{def:Qplusminus}
\end{align} 
where $\overline{T}$ is the horizontal average of temperature. 
Typically, the dimensionless Nusselt number is used to represent the relative strength of convective heat transfer, where the Nusselt number is defined as the ratio of the total heat flux to the heat flux by pure conduction.
For example, using the value of the heat flux at the bottom wall, one can define an instantaneous Nusselt number
\begin{align}
    Nu_-(t) = \frac{\mathcal{Q}_-(t)}{\kappa \Delta T/H} = - \left.\frac{H}{\Delta T}\frac{d\overline{T}}{dy}\right|_{y=-h}.
\label{def:Nu-instantaneous-dimen}
    \end{align}
With the nondimensionalization introduced above, (\ref{def:Nu-instantaneous-dimen}) becomes
\begin{align}
  Nu_-(t) = - \left.\frac{d\overline{T}}{dy}\right|_{y=-1},
\label{def:Nu-instantaneous-nondimen}
\end{align}
and the quantity (\ref{def:Nu-instantaneous-nondimen}) is used herein for plots of Nusselt number as a function of time.
In statistically steady state, the  time-averages of $\mathcal{Q}_{+}$ and $\mathcal{Q}_{-}$ in (\ref{def:Qplusminus}) are equal, and we denote this time average by $\mathcal{Q}$. Then the time-averaged Nusselt number in statistically steady state is given by $Nu = \mathcal{Q}H/(\kappa \Delta T)$.
For $Ra < 1707.8$ before the onset of convection, the value of $Nu$ is unity, since all of the heat transferred is by conduction. For $Ra > 1707.8$, the heat transfer is bolstered by convective effects, and thus $Nu > 1$.

\subsection{Numerical Simulations}

Our numerical simulations were performed using the open-source, MPI-parallelized code Dedalus~\cite{burns2020dedalus, lecoanet2015numerical, couston2018energy},
using a Fourier discretization in the horizontal direction and a Cheybshev discretization in the wall-normal direction. Table~\ref{tab:runs} presents the spatial resolution of selected runs, where $N_{f}$ ($N_c$) is the number of Fourier (Chebyshev) basis functions used in the $x$-direction ($y$-direction). The $3/2$ rule was used for dealiasing. Also tabulated are features of the optimal solution at the given $Ra$ and $Pr$, such as the width $L_{opt}$ of the optimal box, and the optimal Nusselt number $Nu_{opt}$. The convection $Re$ in Table \ref{tab:runs} is computed as $Re = 1/\nu_*$.
\begin{table*}
  \begin{ruledtabular}
    \centering
      \begin{subtable}{.7\linewidth}
        \centering
        \caption{$Pr=7$}
        \def~{\hphantom{0}}
        \begin{tabular}{lccccccc}
          $Ra$ & $N_f \times N_c$ & $Re$ & $L_{opt}$ & $Nu_{opt}$ & $Nu$ & $t_{e}$ & $n_e$ \\ [3pt] \hline 
          $1.1 \times 10^5$ & $1024 \times 128$ & $31$  & $2.59$ & $4.96$ & $4.62$ & $40$ & $62$ \\ 
          $1.2 \times 10^6$ & $1024 \times 128$ & $103$ & $2.02$ & $9.67$ & $7.94$ & $30$ & $60$ \\ 
          $5.1 \times 10^6$  & $1024 \times 128$ & $213$ & $1.66$ & $14.82$ & $11.90$ & $31$ & $26$ \\ 
          $1.1 \times 10^7$    & $1024 \times 256$ & $306$ & $1.39$ & $18.37$ & $14.55$ &$30$ & $27$ \\ 
          $2.8 \times 10^7$  & $2048 \times 256$ & $500$  & $1.14$  & $24.54$ & $19.43$ & $24$ & $33$ \\ 
          $5.5 \times 10^7$  & $2048 \times 256$ & $701$  & $0.99$  & $30.02$ &  $23.34$ & $23$ & $35$ \\
          $1.0 \times 10^8$    & $4096 \times 1024$ & $945$  & $0.87$ & $35.95$ & $27.47$ & $21$ & $10$   \\
        \end{tabular}
      \end{subtable}
      \\
      \begin{subtable}{.7\linewidth}
        \centering
        \caption{$Pr=100$}
        \def~{\hphantom{0}}
        \begin{tabular}{lccccccc}
          $Ra$ & $N_f \times N_c$ & $Re$ & $L_{opt}$ & $Nu_{opt}$ & $Nu$ & $t_{e}$ & $n_{e}$ \\ [3pt] \hline 
          $1.1 \times 10^5 $& $512 \times 128$ & $8$  & $1.40$ & $4.97$ & $4.43$ & $183$ & $16$\\
          $5.5 \times 10^5$& $512 \times 128$ & $18$ & $0.87$ & $7.89$ & $6.71$ & $130$ & $10$ \\
          $1.1 \times 10^6$& $512 \times 128$ & $26$ & $0.65$ & $8.60$ & $7.91$ & $111$ & $11$ \\
          $1.1 \times 10^7$& $1024 \times 256$ & $83$ & $0.40$ & $19.49$ & $14.94$ &$83$ & $10$ \\ 
          %$3 \times 10^7$  & $-$ & $137$ & $15.55$ & $19.49$ & $-$\\ 
          $6.1 \times 10^7$& $2048 \times 256$ & $194$  & $0.27$ & $32.20$ & $23.54$ & $72$ & $11$ \\ 
          $1.2 \times 10^8$& $2048\times 512$ & $250$  & $0.22$ & $39.38$ & $28.10$ & $67$ & $12$ \\ 
          $9.0 \times 10^8$& $4096 \times 512$ & $750$ & $0.14$ & $71.74$ & $47.62$ & $63$ & $10$ \\ 
        \end{tabular}
      \end{subtable}
      \caption{Simulation details: number of Fourier modes $N_f$; number of Chebyshev modes $N_c$; Reynolds number $Re$; width of the optimal box $L_{opt}$; optimal Nusselt number $Nu_{opt}$; eddy turnover time $t_e$; number of eddy turnover times $n_e$; Nusselt number $Nu$.  For comparison to $L_{opt}$, note that $L_{prim}=4.03$.}
      \label{tab:runs}
  \end{ruledtabular}
\end{table*}
Time-integration is accomplished with an adaptive implicit-explicit (IMEX) Runge-Kutta (RK) method. Specifically, we use a $(4,4,3)$ method (four implicit stages, four explicit stages, third order accurate)~\cite{ascher1995implicit}. Moreover, the time-step is adjusted dynamically to ensure that the CFL number is less than unity.
%Adaptive time stepping is carried out using an IMEX Fourth-order Runge-Kutta Scheme ($443$) \cite{ascher1995implicit}, with CFL number $\le 1$. 
%Dealiasing is accomplished with the $3/2$ rule, such that the number of spatial locations in $x$ is $N_x = 3/2 N_{f}$. 
%We monitor the change in $Nu$ with time, and 
To determine when the flow has reached statistically steady state, time averages of $Nu_{-}(t)$ were taken over time windows containing 5 (dimensionless) eddy turnover times $t_e$ defined as
\begin{align}
  \label{eq:te}
  t_e = \frac{4}{\left< v_{rms} \right>_{V}},
  %t_e = \frac{2H}{\left< v_{rms} \right>_{V}},
\end{align}
where $\left< \cdot \right>_V $ represents a volume average and $v_{rms}$ is the temporal root mean square of the vertical velocity~\cite{sakievich2016large}. 
The eddy turnover time is a measure of the time it takes a fluid parcel to traverse the fluid layer height twice.
The flow is considered to be in a statistically steady state once the difference between the successive time averages $Nu$ drops below $1\%$. 
In order to balance high fidelity and computational time, all simulations were run for approximately 10 eddy turnover times after attaining statistically steady state. Mesh refinement was performed for the computations presented in Table~\ref{tab:runs} to ensure that the simulations were well-resolved.

The simulations were initialized from a state of rest, with temperature distribution
\begin{equation}
  %T(x,y, t=0) = 10^{-3} r(x,y) (y - 1)  (y + 1)  
  \Theta(x,y, t=0) = A(x,y) (y - 1)  (y + 1)  
  \label{eqn:initialconditions}
\end{equation}
where $A(x,y) = 0.001r\lr{x,y}$ and $r\lr{x,y}$ are random numbers drawn from a normal distribution. %\lms{Is this $\Theta$?} 
The quadratic form in $y$ satisfies the boundary conditions at the walls, and the noisy amplitude factor allows us to seed the temperature without imposing any horizontal scale. At low $Ra$, we found that the stationary state may exhibit sensitivity to initial conditions. For example, in a domain with aspect ratio $\Gamma = 10$, and for parameters $Ra = 1.1\times10^5$ and $Pr = 7$, we observed two distinct stationary states---one with 4 plumes in the temperature field ($Nu = 4.3$), and the other with 5 plumes ($Nu = 4.6$).   
As will become evident below, for low $Ra$, the horizontal length scale and spacing of the emerging plumes appears to be heavily influenced by the optimal and primary steady solutions. Sensitivity to initial conditions at low $Ra$ has also been discussed in~\cite{venturi2010stochastic,van2011connecting}.

As noted above, the number of thermal structures in the temperature field may be 
influenced by the choice of $\Gamma$. Our objective was to allow the nonlinear interactions, rather than the box size, to select the dominant horizontal scale. In~\cite{stevens2018turbulent} and \cite{bailon2010aspect}, the authors investigated three-dimensional Rayleigh-B\'{e}nard convection at $2 \times 10^7 \le Ra \le 10^9$ and $Pr=1$ at various aspect ratios. They found that the integral length scale and $Nu$ saturate for, respectively, aspect ratio $\Gamma \ge 32$ and $\Gamma \ge 4$. Our 2D studies showed that the scaling of $Nu$ vs.~$Ra$ for $Pr=7$ and $Pr=100$ is largely insensitive to changes in box size beyond $\Gamma =8$. Thus we chose $\Gamma = 10$ to help ensure that the intrinsic dynamics are determining the number and spacing of the plumes in steady state (with some sensitivity for low $Ra$ as mentioned).

We also investigated the domain aspect-ratio dependence of the mean flow (the horizontal average of the horizontal velocity $u(x,y,t)$). We observed that the ratio of mean flow energy and total kinetic energy decreases substantially with increase in $\Gamma$ for $Pr=7$, $Ra = 10^7$ (the energy ratio is $< 0.001 \%$ at $\Gamma = 10$). The authors in~\cite{hartlep2003large} report that the energy contained in the mean flow relative to the total kinetic energy is always less than $0.8\%$ at $\Gamma \approx 10$ for all the cases considered ($Ra \le 10^7$) which includes $Pr=7$ and $Pr=30$. Furthermore, it is observed that the mean flow lowers $Nu$.
Indeed, the primary and optimal solutions have zero mean-flow imposed by symmetry about the $y$-axis. Thus, when searching for the signature of the primary and optimal solutions in the simulation results, the essentially-zero mean flow is an additional motivation for the choice $\Gamma = 10$.   

\begin{figure}[h!]
  \centering
  \includegraphics[width=\textwidth]{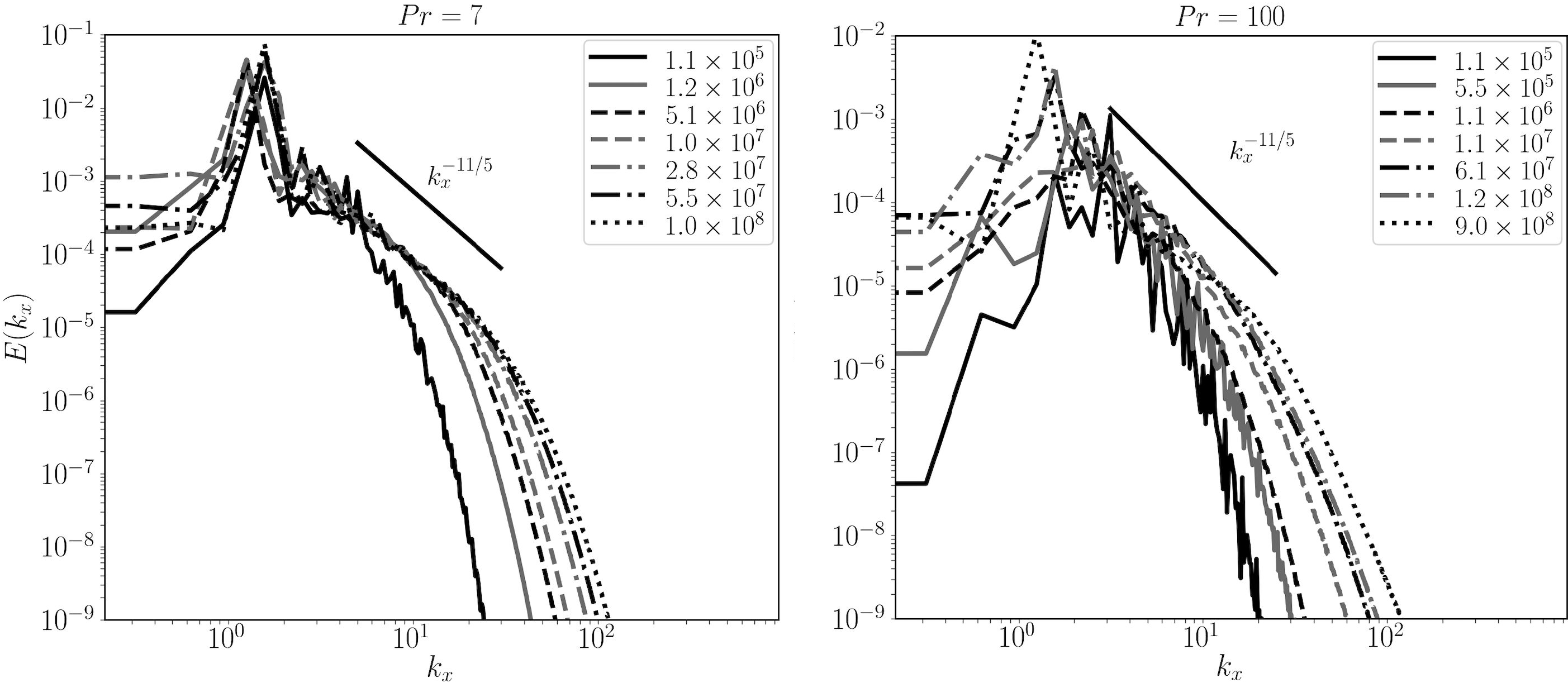}
  \caption{Energy spectra $E(k_x)$ as a function of horizontal wavenumber $k_x$ for $Pr=7$ (left) and $Pr=100$ (right). Spectra are averaged over $y$ and $t$. The values of $Ra$ and corresponding symbols are shown in the insets.}
  \label{fig:KE_k}
\end{figure}

The kinetic energy spectra $E(k_x)$ in statistically steady state are shown in Figure \ref{fig:KE_k}. All the reported runs are broad spectrum and therefore have non-trivial flow dynamics. As $Ra$ increases, the developing inertial range is consistent with the scaling $k_x^{-11/5}$~\cite{mishra2010energy}.  The thermal energy spectra from our simulations also appear to approach the
scaling $k_x^{-7/5}$ reported in previous studies~\cite{chilla1993boundary}.

\section{Methodology}
\label{sec:method}
When considering strategies to detect steady solutions in the time developing simulations, we first explored approaches to search for the signature of the optimal solutions. An obvious strategy is to monitor the global heat transfer and focus on high instantaneous $Nu_{-}\lr{t}$ events. Then, extracting flow fields at individual times corresponding to high $Nu_{-}\lr{t}$, one may inspect these fields for signs of structural similarity with the optimal solutions. Although this method is successful at low $Ra$, it has a number of drawbacks.
First, the global $Nu$ is governed by the combined contributions of {\it all the plumes} present in the field at any given time, and therefore may not be the best indicator of a close match between the optimal solution and a {\it local plume} in the large-aspect-ratio box. Furthermore, we wish to employ a more general technique to detect different steady solutions (optimal, local max, and primary).

\begin{figure}[h!]
  \centering
  \includegraphics[width=\textwidth]{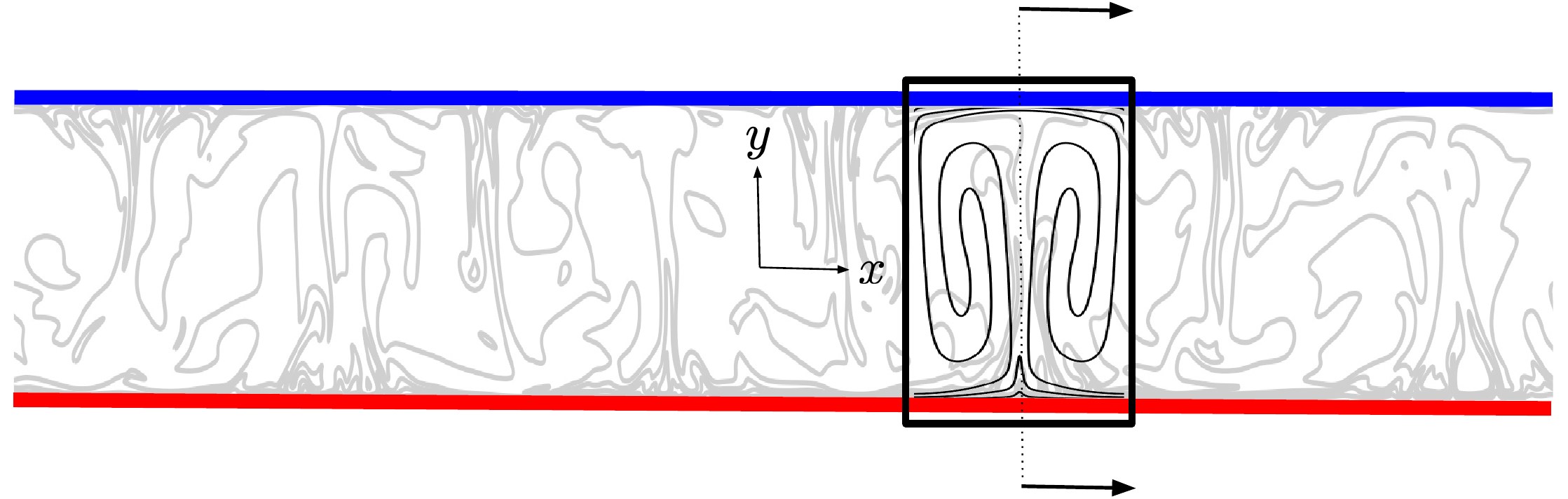}
  \caption{Schematic showing the moving window technique for spatial correlation.}
  \label{fig:moving_window}
\end{figure}

To address these concerns, we developed a windowing technique to search locally for all of the known steady solutions, at all times, keeping $Ra$ and $Pr$ fixed. For fixed time, Figure~\ref{fig:moving_window} illustrates the idea using a schematic of the optimal solution for $Pr=7, Ra = 1.1 \times 10^5$.
By sliding the window across the length of the box, one can compute the spatial correlation between a steady solution and the sub-field centered at $x$, given by
  \begin{align}
    \cos(\theta) = \frac{\left< T_{steady} ,T_{sub} \right>}{||T_{steady}||_{2} ||T_{sub}||_{2}}, 
    \label{eq:align}
  \end{align}
where $\left<\cdot,\cdot\right>$ is an $L^{2}$ inner product and $T_{steady}$ is the temperature field of the steady solution. The notation $T_{sub}$ refers to the temperature field in a portion of the computational domain, which is centered at location $x$ with horizontal extent $2l_{steady}$. Repeating this for every snapshot of the temperature field, one obtains the alignment $\cos\left(\theta\left(x,t\right)\right)$ as a function of position $x$ and time $t$. This spatial correlation data can be used to approximate a probability density function, from which we can quantify the likelihood of convective plumes that are highly correlated with a particular steady solution. For a given time regime, e.g.\ the spin-up period or statistically steady state regime,
we counted the number of sub-fields with correlation $0.8\leq \cos\lr{\theta} \leq 1.0$. We define the incidence parameter $\mathcal{I}$ as
\begin{align}
  \mathcal{I} = \frac{M_{\left[0.8,1.0\right]}}{M_{\text{tot}}} \label{eq:incidence}
\end{align}
where $M_{\left[0.8,1.0\right]}$ is the number of sub-fields within the  specified correlation range, and $M_{\text{tot}}$ is the total number of sub-fields. A few representative values of the incidence parameter $\mathcal{I}$ for the primary solution are given in Table~\ref{tab:stats_prim} and for the optimal solution in Table~\ref{tab:stats}. These numbers will help to inform the discussion in Section \ref{sec:results}.
\begin{table*}
  \begin{ruledtabular}
  \begin{subtable}{0.45\textwidth}
    \centering
    \caption{$Pr=7$ \; (primary) } 
    \begin{tabular}{cLc} 
      $Ra$  & ${\cal I}\;  (\%)$ & $\cos(\theta)_{max}$ \\ [3pt] \hline
      $1.1 \times 10^5 \; (\mathcal{T})$ & $29.5$ & $0.99$ \\ %$0.985$\\
      $1.1 \times 10^5 \; (S_s) $ & $28.5$ & $1.0$ \\ %$0.997$ \\
      $1.1 \times 10^7$ & $6.6$ & $0.90$ \\%$0.901$\\
      $5.5 \times 10^7$ & $6.7$ & $0.84$ %$0.841$
    \end{tabular}
  \end{subtable}
  \hfill 
  \begin{subtable}{0.45\textwidth}
    \caption{$Pr=100$ (primary) }
    \centering
    \begin{tabular}{cLc} 
      $Ra$  & ${\cal I}\;  (\%)$ & $\cos(\theta)_{max}$\\ [3pt] \hline
      $1.1 \times 10^5 \; (Q_s)$  & $20.3$ & $0.94$ \\ %$0.944$\\ 
      $1.1 \times 10^5 \; (\mathcal{T})$ & $28.0$ & $0.99$ \\ %$0.994$\\
      $1.1 \times 10^5 \; (Q_p) $ & $30.0$ & $0.99$ \\ %$0.990$ \\
      $1.1 \times 10^7$ & $42.5$ & $0.87$ \\ %$0.874$\\
    \end{tabular}
  \end{subtable}
  \end{ruledtabular}
  \caption{Comparison of statistical features of the PDFs for correlation of sub-fields with the primary solution. The incidence $\mathcal{I}$ is the ratio of the number of sub-fields with correlation in the range $[0.8, 1]$ and the total number of sub-fields (see~\eqref{eq:incidence}). For $Pr=7$, $Ra = 1.1 \times 10^5$, the symbols refer to time regimes in Figure \ref{fig:Nu_a1E5_7}: transient $\mathcal T$ for $t < 1500$; statistically steady $S_s$ for $t > 1750$. Similarly, for $Pr = 100$, $Ra = 1.1\times 10^5$, the symbols refer to Figure \ref{fig:Nu_t_1E5_100}: quasi-steady $Q_s$ for $200 < t < 5000$; transient $\mathcal T$ for $5000 < t < 23,000 $; quasi-periodic $Q_p$ for $ t > 23,000$. All incidence values are measured in statistically steady state with a minimum of 10 eddy turnover times.}
  \label{tab:stats_prim}
\end{table*}
\begin{table*}
  \begin{ruledtabular}
  \begin{subtable}{0.45\textwidth}
    \centering
    \caption{$Pr=7$ (optimal)} 
    \begin{tabular}{cLc} 
      $Ra$  & ${\cal I} \; (\%)$ & $\cos(\theta)_{max}$ \\ [3pt] \hline
      $1.1 \times 10^5 \; (\mathcal{T}) $ & $27.6$ & $1.0$ \\ %$0.996$\\
      $1.1 \times 10^5 \; (S_s) $ & $29.9$ & $0.97$ \\ %$0.972$\\ 
      $1.2 \times 10^6$ & $5.27$ & $0.90$ \\ %$0.900$\\
      %$5.1 \times 10^6$ & $0.07$ & $0.844$ \\
      $1.1 \times 10^7$ & $0.02$ & $0.82$ \\ %$0.823$\\
      %$2.8 \times 10^7$ & $0.05$ & $0.847$\\
      %$5.5 \times 10^7$ & $0$ & $0.760$\\
      %$1.0 \times 10^8$  & $0.01$ & $0.815$
      %\\
    \end{tabular}
  \end{subtable}
  \hfill 
  \begin{subtable}{0.45\textwidth}
    \caption{$Pr=100$ (optimal)}
    \centering
    \begin{tabular}{cLc} 
      $Ra$  & ${\cal I} \; (\%)$ & $\cos(\theta)_{max}$\\ [3pt] \hline
      $1.1 \times 10^5 \; (Q_s)$ & $28.7$  & $1.0$ \\ % $0.997$ \\
      $1.1 \times 10^5 \; (\mathcal{T})$ & $40.7$  & $0.99$ \\ % $0.994$ \\
      $1.1 \times 10^5 \; (Q_p)$ & $27.9$  & $0.94$ \\ % $0.938$ \\
      %$5.5 \times 10^5$ & $23.6$ & $0.922$ \\
      %$1.1 \times 10^6$ & $15.9$ & $0.935$ \\
      $1.1 \times 10^7$ & $1.0$ & $0.89$ \\ % $0.885$ \\
      %$6.1 \times 10^7$ & $0.02$ & $0.853$ \\
      %$1.2 \times 10^8$ & $0$ & $0.859$\\
      %$9.0 \times 10^8$ & $0$ & $0.763$
      %\\
    \end{tabular}
  \end{subtable}
  \end{ruledtabular}
  \caption{Comparison of statistical features of the PDFs for correlation of sub-fields with the optimal solution.
  All symbols have the same meaning as explained in Table \ref{tab:stats_prim}.}
  \label{tab:stats}
\end{table*}

The windowing technique allows us to identify the influence of multiple steady states on the structure and statistics of transitional flows, especially with respect to horizontal scale selection by nonlinear interactions. For this first study, beyond our lowest $Ra \approx 10^5$, we focus on the primary and optimal solutions.  
We observe that several of their features can remain intact 
for $10^{7}\lesssim Ra \lesssim 10^{8}$ at both $Pr$ considered in this work.  It is remarkable that their signatures endure, since turbulence significantly disrupts their wall-to-wall nature, along with their upright plumes, and in some cases, their delicate interior structures.  
%For high enough $Ra$, turbulence ultimately destroys the wall-to-wall nature of the steady solutions, along with their upright plumes, and their delicate interior structures.  
Once highly-correlated sub-fields have been identified, singular value decomposition~\cite{trefethen1997numerical} is used for more detailed comparison between steady solutions and instantaneous simulation sub-fields. We find that the first two SVD modes are associated with the boundary layer and plume structures, respectively, and provide a quantitative means to separately assess agreement for these two structural features.

\section{Results}\label{sec:results}

%For points of comparison, 
To set the stage, we begin this section by reviewing some relevant literature describing the dynamics of similar $Ra-Pr$ regimes studied herein. In \cite{paul2011bifurcations}, the authors employed a low dimensional model using the most energetic modes from direct numerical simulations to study 2D flow regimes at $Pr = 6.8$.
According to \cite{paul2011bifurcations}, the flow is chaotic for $r_c > 48.4$, where $r_c = Ra/Ra_c$ and $Ra_c \approx 1708$. Experimental studies in \cite{gollub1980many} investigated the many routes the flow takes to reach a turbulent state, and the dependence of these routes on aspect ratio $\Gamma$, $Pr$, and the presence of a mean flow. Their results for $\Gamma = 3.5$ at $Pr=5$ show that transition to a non-periodic state occurs for $r_c > 50$. For $Pr = 7$, the lowest value of $r_c$ investigated in our simulations is $r_c=64.4$, expected to be well above the threshold for chaos by comparison to 
\cite{paul2011bifurcations,gollub1980many}. In the experimental studies \cite{xia2002heat} with $\Gamma = 1$ at $4\le Pr \le 1350$, the flows with $Ra \ge 2 \times 10^7$ have been described as being turbulent. 

We also conducted studies of our own to classify the nature of our simulation fields. For example, we examined the time traces of vertical velocity and temperature, collected from probes placed in the bulk and the boundary layer.  With the exception of the lowest $Ra = 1.1 \times 10^5$ for $Pr = 100$, all cases exhibited spectra with a broad wavenumber and frequency distribution in statistically steady state (see Figure~\ref{fig:KE_k}).

We next present our results, organized into two separate sections for $Pr = 7$ and $Pr=100$ (Section~\ref{subsec:Pr7new} and~\ref{subsec:Pr100new}, respectively). As we will demonstrate, primary and optimal structures are readily observed in temperature fields corresponding to transitional values of $Ra$. Visualizations and data analysis of the simulations for higher $Ra$ suggest that a signature of the primary solution persists for $Pr = 7$, consistent with agreement for $Nu$ vs.\ $Ra$ scaling seen in Figure~\ref{fig:Nu-Ra-AR10}. The optimal solution for higher $Ra$ is easier to detect in the case of $Pr=100$, perhaps in part because of its simpler structure compared to the optimal solution for $Pr=7$ (see Figure~\ref{fig:prim_opt_comparison}).

\subsection{$Pr = 7$}
\label{subsec:Pr7new}

\subsubsection{Visualizations and correlation data}
\label{subsubsec:visPr7}

\begin{figure} [h!]
    \centering
    \includegraphics[width=0.85\textwidth]{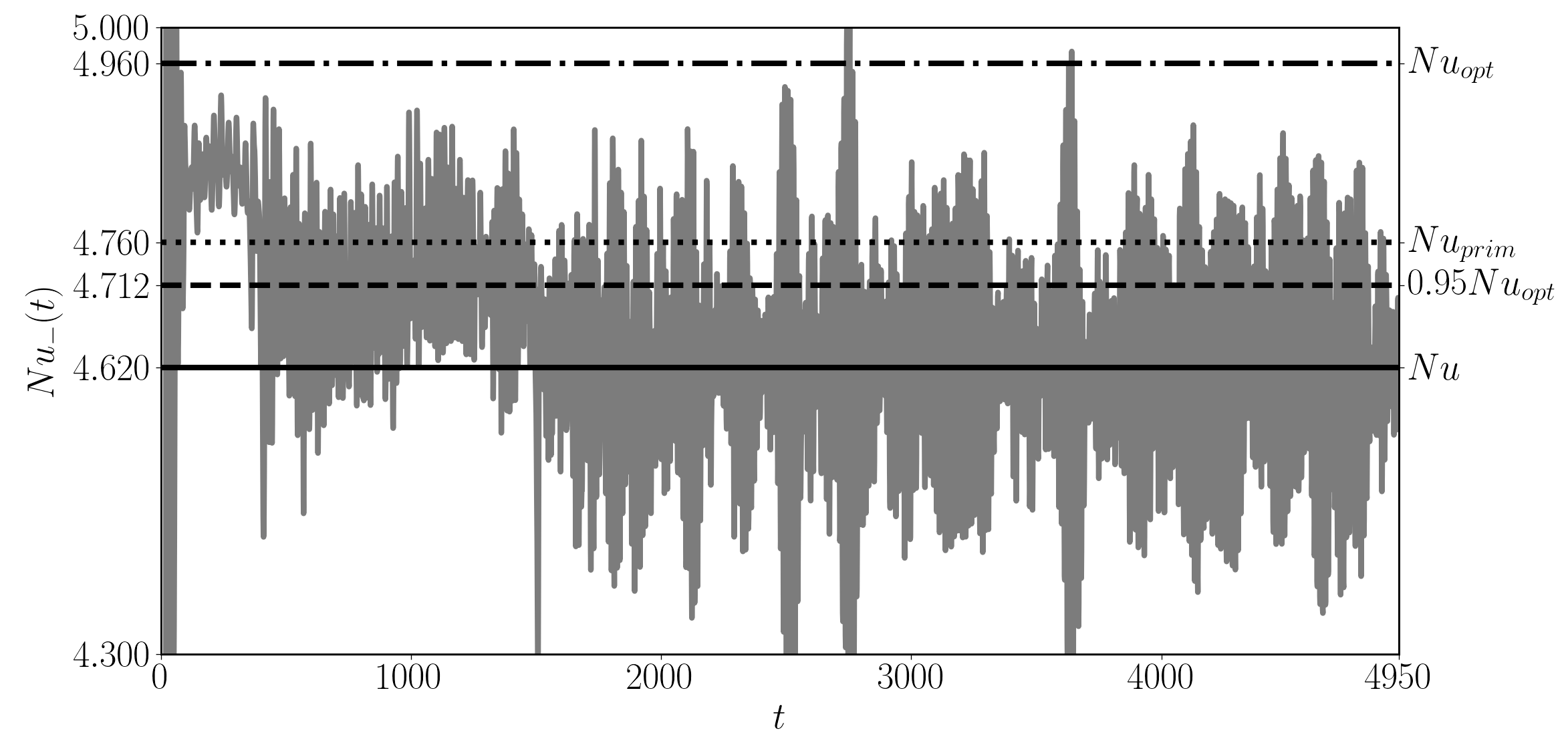}
    \caption{Evolution of $Nu_{-}\lr{t}$ for $Ra= 1.1 \times 10^5$, $Pr=7$. A chaotic statistically steady state is reached for $t > 1750$. 
   % Horizontal lines indicate: $Nu$ for the simulation in statistically steady state (solid): $Nu_{opt}$ for the optimal solution  (dash-dot) and $95\%$ of $Nu_{opt}$ (dash); $Nu_{prim}$ for the primary solution (dot).
   }
\label{fig:Nu_a1E5_7}
\end{figure}

We first present results for transitional $Ra=1.1 \times 10^5$ ($Re = 31$ and $r_c = 64.4$), starting with the time evolution of the instantaneous Nusselt number in Figure~\ref{fig:Nu_a1E5_7}.  After a significant transient period $t < 1500$, the system eventually settles to a chaotic statistically steady state for $t > 1750$.
In statistically steady state, the Nusselt number is approximately $Nu=4.62$, but one can see large fluctuations in the instantaneous Nusselt number, frequently attaining values greater than $95\%$ of the optimal value $Nu_{opt} = 4.96$. The $Nu$ vs.\ $Ra$ scaling in agreement with the primary solution (Figure~\ref{fig:Nu-Ra-AR10}), together with the near-optimal, values of $Nu_{-}\lr{t}$ (Figure~\ref{fig:Nu_a1E5_7}), suggest that both primary and optimal solutions may be influencing the structure of the boundary layer in this transitional flow.  
Hence we investigate this possibility using our windowing technique, followed later by singular value decomposition of the highly correlated fields for closer inspection of the boundary layers.
Note that the incidence parameters for the primary and optimal solutions are quite close to each other, and greater than $20\%$, in both the transient and statistically steady regimes (see Tables \ref{tab:stats_prim} and \ref{tab:stats}).    

\begin{figure*}[h!]
  \centering
  \includegraphics[width=\textwidth]{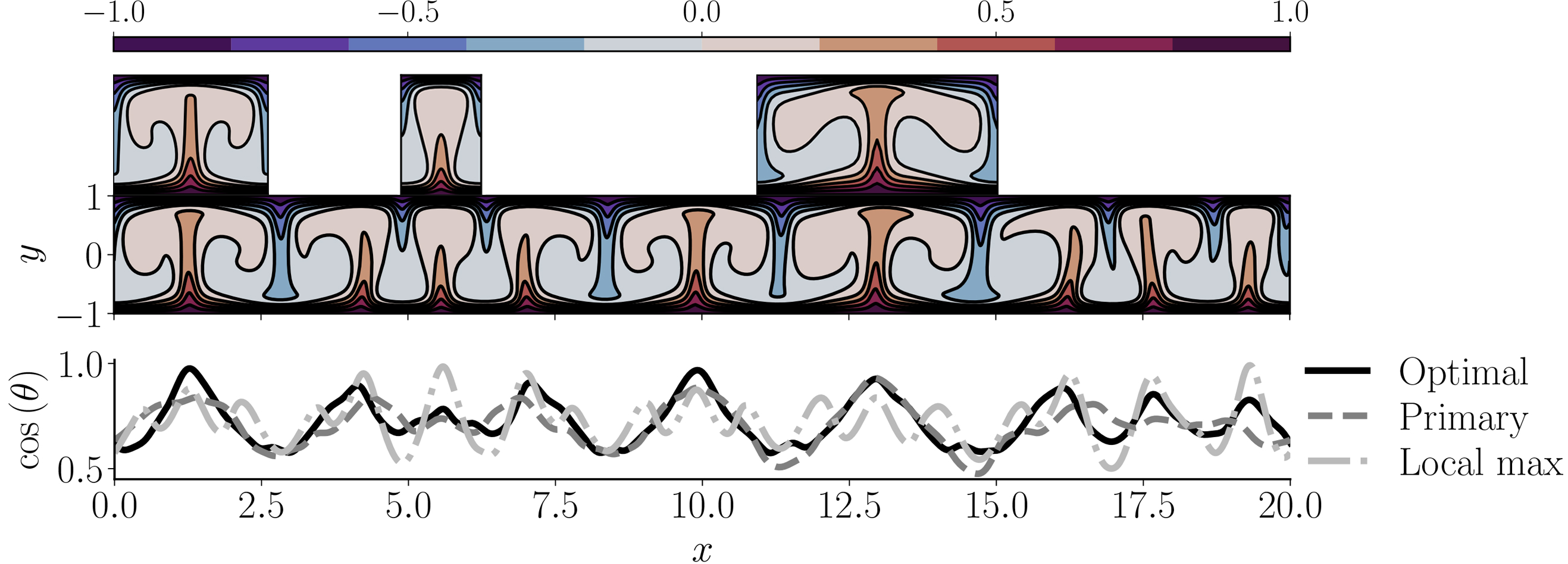}
  \caption{Temperature fields and correlation data for $Ra = 1.1 \times 10^5$ and $Pr=7$. 
  Top panel: (left to right) optimal, local maximal and primary solutions;  Middle panel:
  simulation snapshot at an early time $t = 240$ during spin-up; Bottom panel: spatial correlation $\cos(\theta \left( x \right))$ for the steady solutions and the simulation sub-fields located directly underneath.}
  \label{fig:1E5_Pr7_t240}
\end{figure*}

Figure~\ref{fig:1E5_Pr7_t240} (middle panel) displays an instantaneous snapshot of the temperature field in the simulation with $Ra=1.1\times 10^{5}$. The snapshot corresponds to the early time $t=240$, when different horizontal length scales are emerging from nonlinear interactions, quite clearly corresponding to the scales associated with the primary, optimal and local maximal heat transport solutions (top panel).
The situation is similar to Figure~\ref{fig:1E5_100_2} for $Pr=100$ at the same $Ra = 1.1 \times 10^5$ (though the values of $Re$ are different). In Figure~\ref{fig:1E5_Pr7_t240}, the bottom panel shows two prominent peaks in the correlation function defined by~\eqref{eq:align} for the optimal solution, roughly located at $x \approx 1.3$ and $x \approx 10$. Centered at these locations, there is a temperature plume in the time-developing simulation with visual similarity to the optimal solution, especially with respect to horizontal scale and overall shape (see the top panel for a comparison centered at $x \approx 1.3$).  The analogous correlation function for the second maximal heat transport solution has six peaks, and again, one observes overall structural similarity between the steady solution and updrafts in the simulation field. At this early time ($t = 240$), the primary solution is highly correlated with only one temperature plume located at $x\approx 13$.  These observations %from Figure~\ref{fig:1E5_Pr7_t240} 
are consistent with the high values, close to optimal, achieved by the instantaneous Nusselt number during early spin-up times $t < 400$.
\begin{figure*}[h!]
  \centering
  \includegraphics[width=\textwidth]{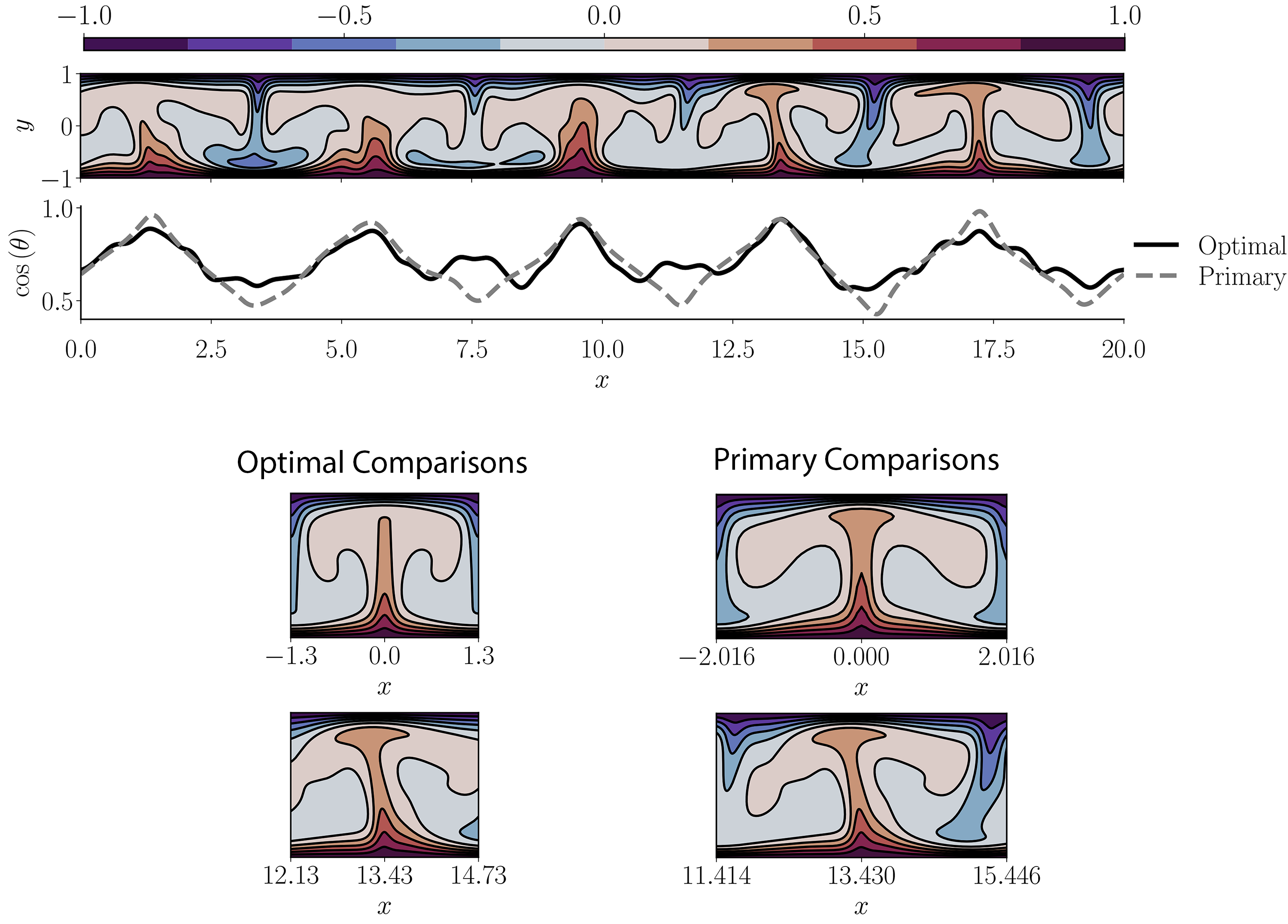}
  \caption{Temperature fields and correlation data for $Ra = 1.1 \times 10^5$ and $Pr=7$. Top panel: simulation snapshot at $t = 1827$ in the chaotic, statistically steady regime; Middle panel: spatial correlation $\cos(\theta \left( x \right))$ for the solutions and the simulation sub-fields; Bottom left column: comparison of the optimal solution to the plume centered at $x=13.4$ in the computational domain; Bottom right column: comparison of primary solution to the same plume centered at $x=13.4$.}
  \label{fig:spatial_correlation_st_opt}
\end{figure*} 

\begin{figure}[h!]
  \centering
  \includegraphics[width=0.9\textwidth]{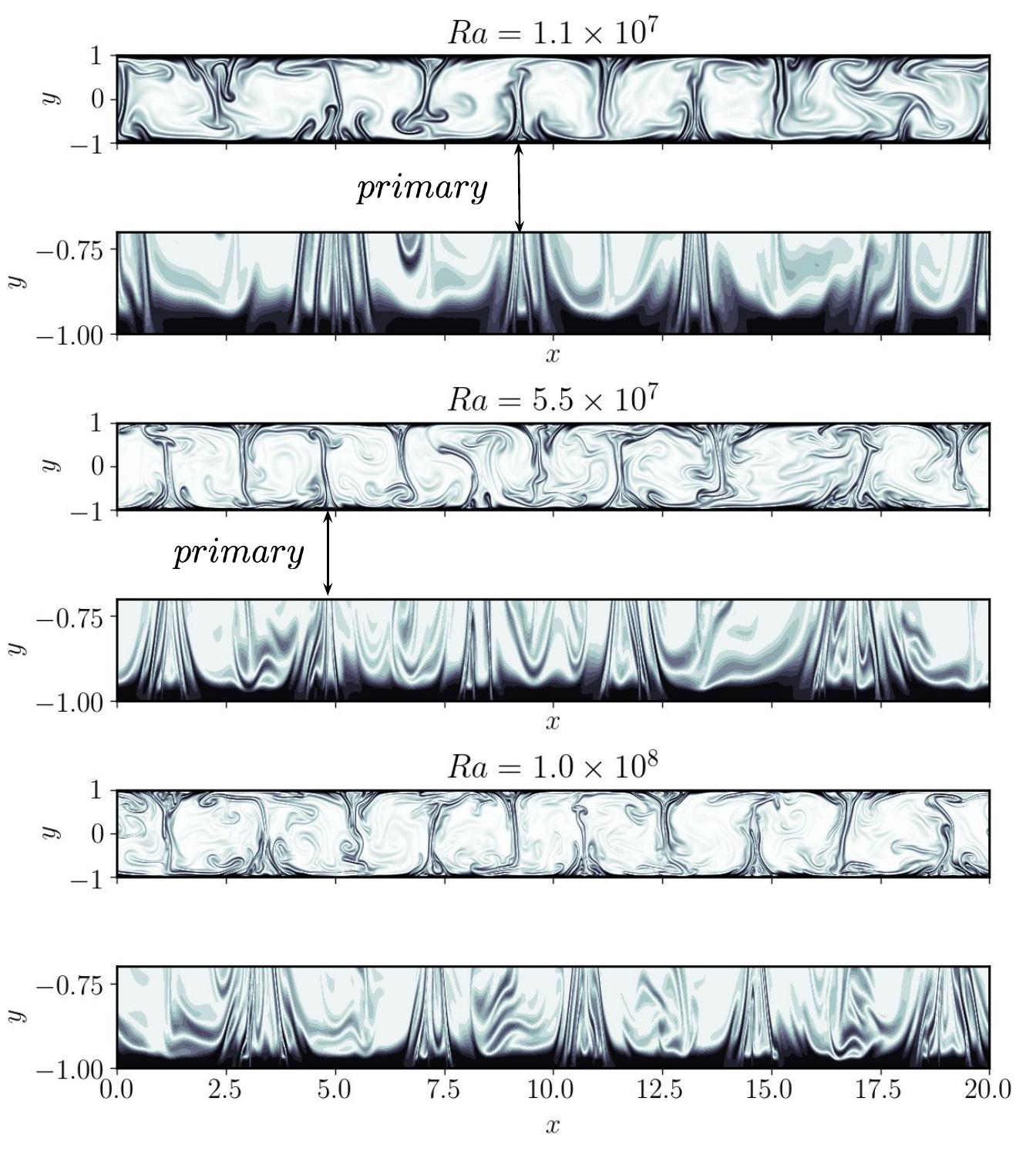}
  \caption{Snapshots of temperature at $Pr = 7$.  For three values of $Ra$, there is a full temperature field (top) and a zoom on the lower boundary (underneath).  The $Ra=1.1 \times 10^{7}$ and $Ra=5.5\times 10^{7}$ plots are marked with an arrow pointing to a turbulent plume that is well correlated with the primary solution. The values of the parameter $\gamma$ in (\ref{eqn:schlieren}) are $\gamma = 0.5$ in full fields and $\gamma=0.2$ in the zoom on the lower boundary.}
  \label{fig:T_7}
\end{figure}

As the simulation progresses, the system eventually settles into a statistically steady state with representative temperature field shown on the top panel of Figure~\ref{fig:spatial_correlation_st_opt}. At the time $t = 1827$ shown in the figure, the correlation functions for the optimal and primary solutions (middle panel) both show five distinct peaks corresponding to the temperature updrafts.  
The bottom left column (bottom right column) compares the optimal (primary) solution and a single plume of the turbulent snapshot centered at $x=13.4$.  Note that the size of the alignment window is different for the same plume because the window is determined by the horizontal length scale of the optimal or primary solution.  The correlation functions show a slightly better match between the primary solution and instantaneous updrafts.  From Tables \ref{tab:stats_prim} and \ref{tab:stats}, we also observe that the incidence ${\cal I}$ for the primary solution is roughly $20\%$ higher than the incidence for the optimal solution, after the flow has transitioned to statistically steady state at $Ra = 1.1 \times 10^5$.
The information from the correlation data reflects, in part, the tendency of the nonlinear interactions to select the horizontal scale of the primary solution.  As also seen from 
Tables \ref{tab:stats_prim} and \ref{tab:stats}, while both incidence values decrease for increasing $Ra$, the gap between the primary solution incidence and the optimal solution incidence grows.  That gap is consistent with the $Nu$ vs.\ $Ra$ data in Figure \ref{fig:Nu-Ra-AR10}, and also with the spacing of large-scale horizontal plumes as visualized in Figure \ref{fig:T_7}. 

We note that the correlation $\cos({\theta})$ weighs similarities in regions of low and high temperature ($T \approx \pm 1$) more than in the regions with temperatures near zero. 
Since the same boundary conditions are imposed for the steady solutions and the time-developing flows, necessarily leading to top and bottom boundary layers, the correlations rarely go below the value $\cos(\theta) = 0.4$. Furthermore, the temperature contours of highly correlated structures picked out by the windowing algorithm match well with the thermal structures in the top and bottom boundary layers, rather than in the bulk where the temperature distribution is close to zero.  As mentioned above, this must be kept in mind for higher values of $Ra$, when fluctuations become more vigorous leading to fully developed turbulence.
At high $Ra$, the plumes will not be upright, and interior structures of the steady solutions become fractured, such as the coiling arms of the primary solution (see Figure \ref{fig:Ra1E7_Pr7}).  Indeed, the incidence parameters shown in Tables~\ref{tab:stats_prim} and~\ref{tab:stats} decrease accordingly with increasing $Ra$. Nevertheless, for values of $Ra$ up to $Ra = 1.0 \times 10^8$, Figure~\ref{fig:T_7} illustrates that very thin `plumelets' converge in the boundary layer to produce larger-scale updrafts that are nearly wall-to-wall, with spacing that appears to be approximately given by 
%closely related to 
the horizontal scale of the primary solution.  Figure~\ref{fig:T_7} is a Schlieren-type plot of the function
\begin{align}
    T_{Schlieren} = \exp\lr{-\gamma\dfrac{\left\|\nabla T\right\|}{\left\|\nabla T\right\|_{\text{max}}}},
\label{eqn:schlieren}
\end{align}
where $\left\|\nabla T\right\|$ is the magnitude of the temperature gradient at each point in the domain, $\left\|\nabla T\right\|_{\text{max}}$ is the maximum temperature gradient magnitude over the domain, and $\gamma$ is a scaling parameter.  
The values of $\gamma$ are different in the bulk and boundary layers and specific values are reported along with the accompanying figures. The exponential function helps to bring out flow features~\cite{quirk1997contribution} that may otherwise be washed out in the visualization. Counting the number of large-scale updrafts in our statistically steady flows for $10^5 < Ra < 10^8$, one consistently arrives at the number (approximately) 5 in the domain of length $L=20$. Since the width of the primary solution is $2l_p = 4.03$, this strongly suggests  that the horizontal scale selected by the disordered flow is determined by the primary solution, at least in this range of $Ra.$ 

\subsubsection{Singular Value Decomposition of Steady Solutions and Simulation Sub-fields}
\label{sec:svd_Pr7}

Here we further scrutinize the dominant features of simulation sub-fields selected by the windowing algorithm for a more detailed comparison to the steady solutions, either primary or optimal. Standard singular value decomposition (SVD) \cite{trefethen1997numerical} is used to analyze the temperature field
of the steady solutions and the simulation sub-fields. 
To select a specific sub-field for the SVD analysis, we first isolate sub-fields with correlation values greater than or equal to $0.9\cos(\theta)_{max}$. These sub-fields are filtered by averaging the temperature profiles in the range $y=[-1,0.5]$,
and then selecting those sub-fields with a peak in the averaged temperature in the range $x=[-l/4, l/4]$, where the length of the sub-field is $2l$. The specific instances for SVD are chosen blindly from this subset. Note that this algorithm can select sub-fields with lower correlation value than those counted in the incidence parameter~\eqref{eq:incidence}.

\begin{figure}[h!]
    \centering
    \includegraphics[width=\textwidth]{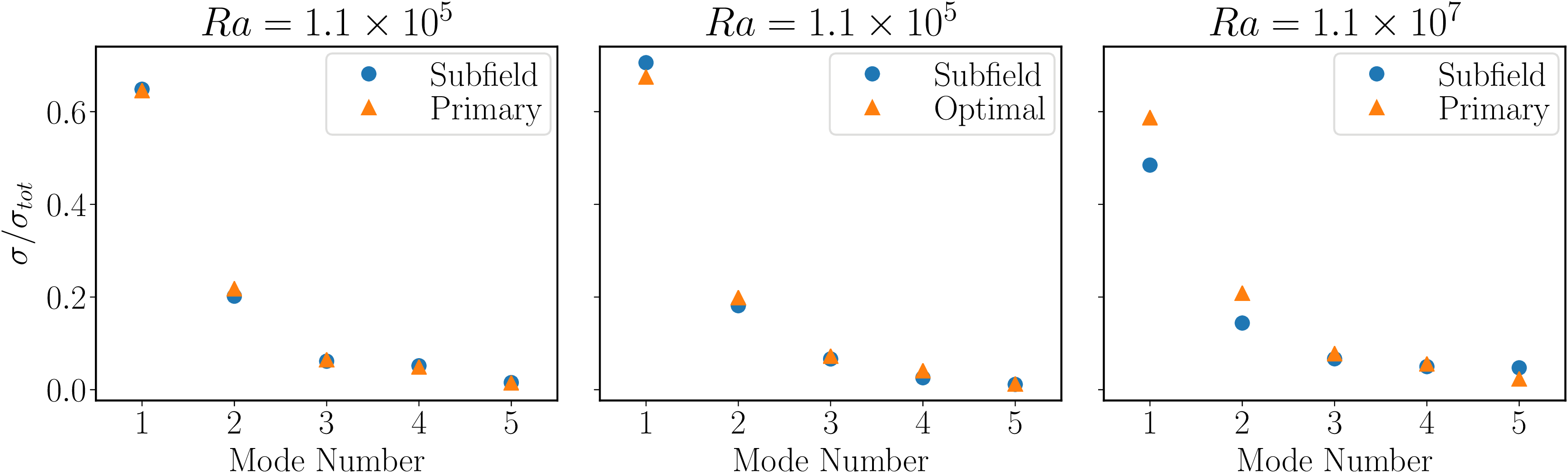}
    \caption{Normalized singular values $\sigma/\sigma_{tot}$, where $\sigma_{tot}$ is the sum of all the singular values. 
    Comparisons are for a simulation sub-box and: (left) the primary solution at $Ra=1.1\times 10^{5}$; (middle) the optimal solution at $Ra=1.1\times 10^{5}$; (right) the primary solution at $Ra=1.1\times 10^{7}$.
    The windows corresponding to the simulations are extracted from times in the statistically steady regime.}
    \label{fig:singular_values_Pr7}
\end{figure}

\begin{figure}[h!]
    \centering
    \includegraphics[width=\textwidth]{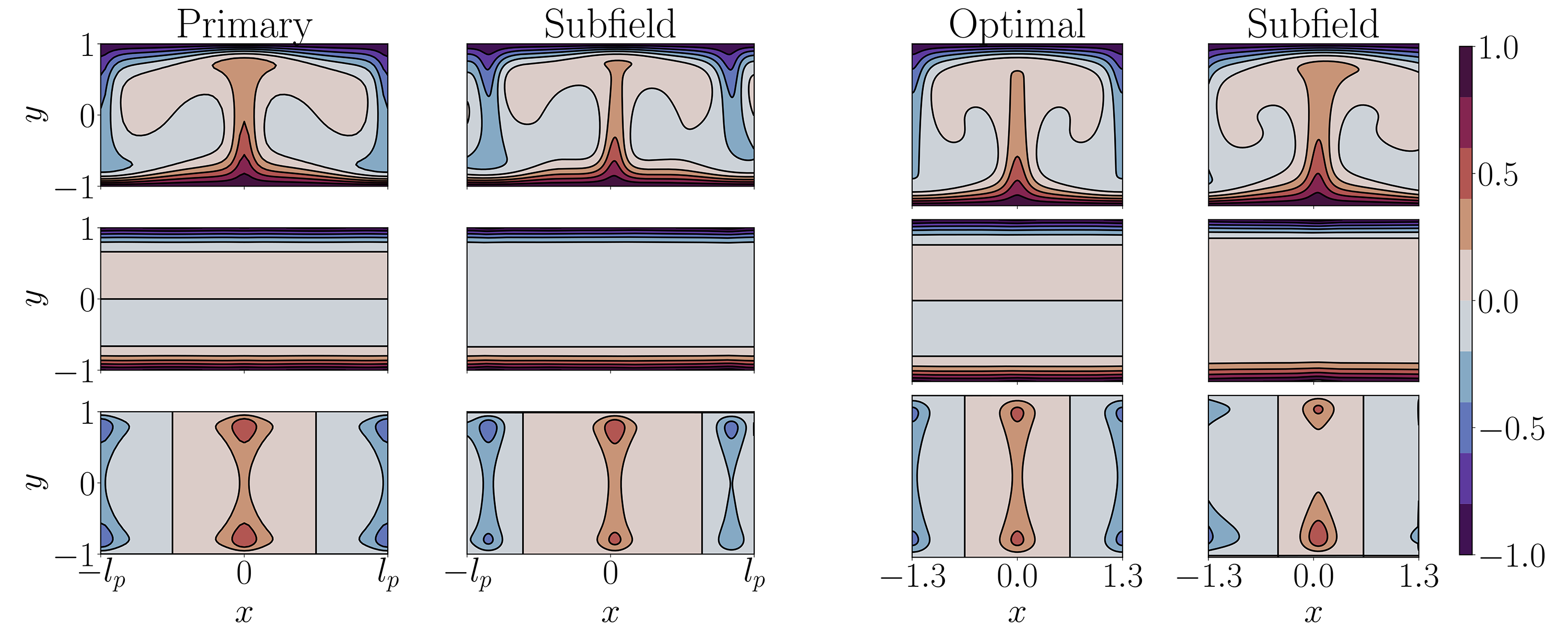}
    \caption{Comparisons between steady solutions and highly-correlated simulation sub-fields at $Ra=1.1 \times 10^5$ and $Pr=7$.  The left two columns (right two columns) compare the primary solution (optimal) solution with a simulation window.  The color bar is shared, and the simulation snapshots are taken from the statistically steady regime.}
    \label{fig:Ra1E5_Pr7_prim-opt}
\end{figure}

\begin{figure}[h!]
    \centering
    \includegraphics[width=0.7\textwidth]{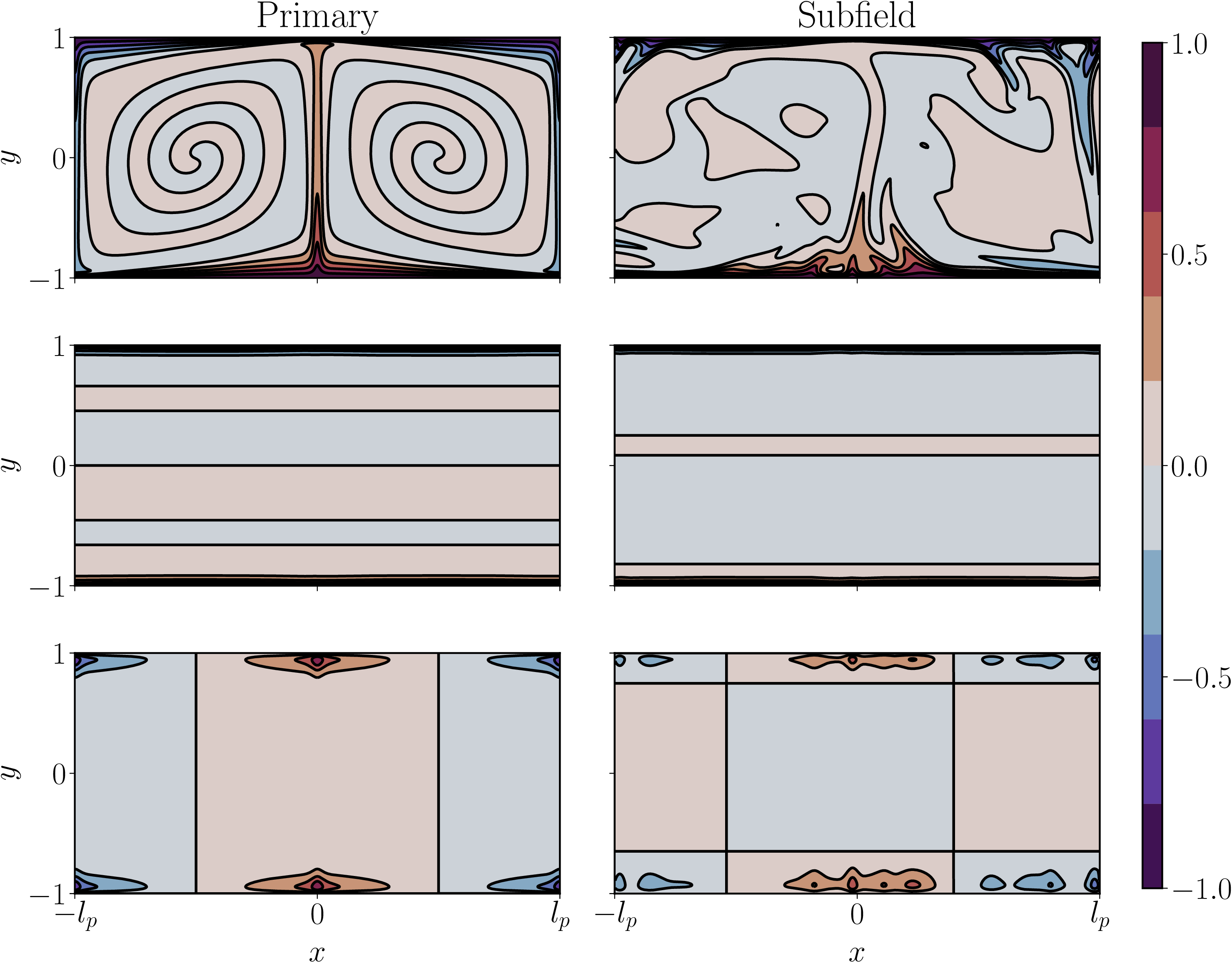}
    \caption{Comparison of primary solution and a highly-correlated simulation sub-field at $Ra=1.1 \times 10^7$ and $Pr=7$.  The color bar is shared, and the simulation snapshot is taken from the statistically steady regime.}
    \label{fig:Ra1E7_Pr7}
\end{figure}

In Figure \ref{fig:singular_values_Pr7}, we compare the singular values of the first 5 SVD modes for three cases: $Ra = 1.1 \times 10^5$ (primary and optimal) and $Ra = 1.1 \times 10^7$ (primary).  In each case, the simulation sub-fields  correspond to the statistically steady time regime.
%and are selected using the algorithm described above.
%In each case, 
%the singular values of the steady solutions are compared to those of a highly correlated simulation window, at a time when the simulation has reached statistically steady state.  
As is traditional, one may interpret each singular value $\sigma$ as an `energy,' and the sum over all values $\sigma_{tot}$ as the total energy.  The ratio 
$\sigma_n/\sigma_{tot}$ is interpreted as the percent of energy in any given SVD mode with number $n$.
As can be seen the figure, roughly $90\%$ of the total energy is contained in SVD modes 1 and 2. 
For the same three cases,
Figures~\ref{fig:Ra1E5_Pr7_prim-opt}-\ref{fig:Ra1E7_Pr7} show that these two SVD modes correspond to, respectively, the boundary layer and updraft-downdraft structures of both steady solutions and simulation sub-fields. The SVD analysis provides an additional quantitative comparison of these structures, adding to the $Nu$ vs.\ $Ra$ information in Figure~\ref{fig:Nu-Ra-AR10}, and the correlation data provided in the previous section.

\begin{table*}
\begin{ruledtabular}
\centering
\begin{tabular}{ccccccc} 
$Ra$  & $\left(\frac{\sigma_1}{\sigma_{tot}}\right)_{s}$ & $\left(\frac{\sigma_1}{\sigma_{tot}}\right)_{sub}$ & Relative Error & $\left(\frac{\sigma_2}{\sigma_{tot}}\right)_{s}$ & $\left(\frac{\sigma_2}{\sigma_{tot}}\right)_{sub}$ & Relative Error \\ [3pt] \hline
$1.1 \times 10^5$ (primary) & $0.645$ & $0.648$ & $0.510\%$ & $0.217$ & $0.202$ & $7.23$\% \\
$1.1 \times 10^5$ (optimal)& $0.674$ & $0.706$ & $4.6\%$ & $0.198$ & $0.182$ & $8.62\%$ \\
$1.1 \times 10^7$ (primary) & $0.587$ & $0.485$ & $17.4\%$ & $0.208$ & $0.144$ & $30.8\%$
\end{tabular}
\end{ruledtabular}
\caption{Comparison of relative energy content in SVD modes 1 and 2 for the primary/optimal ($s$) solutions and the sub-fields ($sub$) of the simulations (data and definitions as in Figure \ref{fig:singular_values_Pr7}).}
\label{tab:stats_svd}
\end{table*}

For $Ra = 1.1 \times 10^5$, Figure~\ref{fig:Ra1E5_Pr7_prim-opt} shows contour plots of the 1st and 2nd SVD modes of the primary and optimal solutions, compared with highly-correlated simulation sub-fields.   
The left two columns 
present the comparison of the primary solution to a simulation sub-field with correlation %$\cos(\theta) = 0.927$.  
$\cos(\theta) = 0.93$. For the first mode corresponding to the boundary layer, the absolute error over the domain ranges between $\approx 3\times 10^{-7}$ in the boundary layer to $\approx 10^{-1}$ in the middle of the domain. Near the boundary layer, the first SVD mode of the primary solution is in remarkably good agreement with the first SVD mode of the turbulent plume. 
Contour plots of the absolute error are given in Figure~\ref{fig:Ra100K_Pr7_Prim-Turb_errs} of Appendix~\ref{app:err_plots}. We also note that a typical value of the absolute error is $2\times 10^{-4}$ in the boundary layer [$(x,y) = (0,-0.779)$] and $3\times 10^{-2}$ in the bulk [$(x,y)=(0, -0.00612)$]. 
The second mode corresponding to the updraft-downdraft exhibits a similarly good agreement (Figure~\ref{fig:Ra1E5_Pr7_prim-opt} and Figure~\ref{fig:Ra100K_Pr7_Prim-Turb_errs} in Appendix~\ref{app:err_plots}). 
Comparing contours of the same levels, SVD mode 1 reveals an extremely close match between boundary layer thicknesses for the primary solution and the simulation window.  To make this comparison quantitative, Table \ref{tab:stats_svd} presents the normalized singular values for SVD mode 1, where the relative error reflects agreement between the boundary layer heights. Notice the small relative error value of $0.5\%$ for SVD mode 1 analysis of the primary solution at $Ra = 1.1 \times 10^5$.

The right two columns of Figure~\ref{fig:Ra1E5_Pr7_prim-opt} present the comparison between SVD modes for the optimal solution and a simulation sub-field with correlation %$\cos\lr{\theta} = 0.943$. 
$\cos\lr{\theta} = 0.94$. Once again, the first SVD mode corresponds to the boundary layer, however, the agreement is not as close as observed for the primary solution comparison. 
Figure~\ref{fig:Ra100K_Pr7_Opt-Turb_errs} in Appendix~\ref{app:err_plots} provides contours of the absolute error for this case. The boundary layer in the optimal solution is smaller than the boundary layer of its companion sub-field, and Table \ref{tab:stats_svd} gives a relative error of $4.6\%$. Since $Nu$ scales with the reciprocal of the boundary layer thickness, this observation is consistent with Figure~\ref{fig:Nu-Ra-AR10}, in which $Nu_{\text{opt}}$ provides an upper bound on $Nu$ for the primary and turbulent solutions. As an upper bound on heat transport, the optimal steady solution necessarily has thinner boundary layer than the average boundary layer thickness in the simulation domain.
The rather tight nature of the bound, especially at transitional values of $Ra$, is captured by the SVD mode 1 analysis of highly correlated simulation windows with the optimal length scale.

We finish this section with Figure \ref{fig:Ra1E7_Pr7}, which illustrates how the primary solution persists at higher $Ra = 1.1 \times 10^7$, albeit with unstable boundary layer, and only distant remnants of the coiling arms. 
The location of the simulation sub-field is shown by the arrow in the 3rd panel of Figure \ref{fig:T_7}.  As can be seen even more clearly from Figure \ref{fig:Ra1E7_Pr7} than from Figure \ref{fig:T_7}, two important features of the primary solution remain intact: 
a boundary layer (1st SVD mode), and the horizontal scale associated with an updraft-downdraft pair (2nd SVD mode). In place of the plume-like structures associated with the 2nd mode at $Ra = 1.1 \times 10^5$ (Figure \ref{fig:Ra1E5_Pr7_prim-opt}), now the 2nd mode could be described as the `hotspots' (`cold pools') at the upwellings (downwellings) of temperature.

\begin{figure}[h!]
  \centering
  \includegraphics[width=0.8\textwidth]{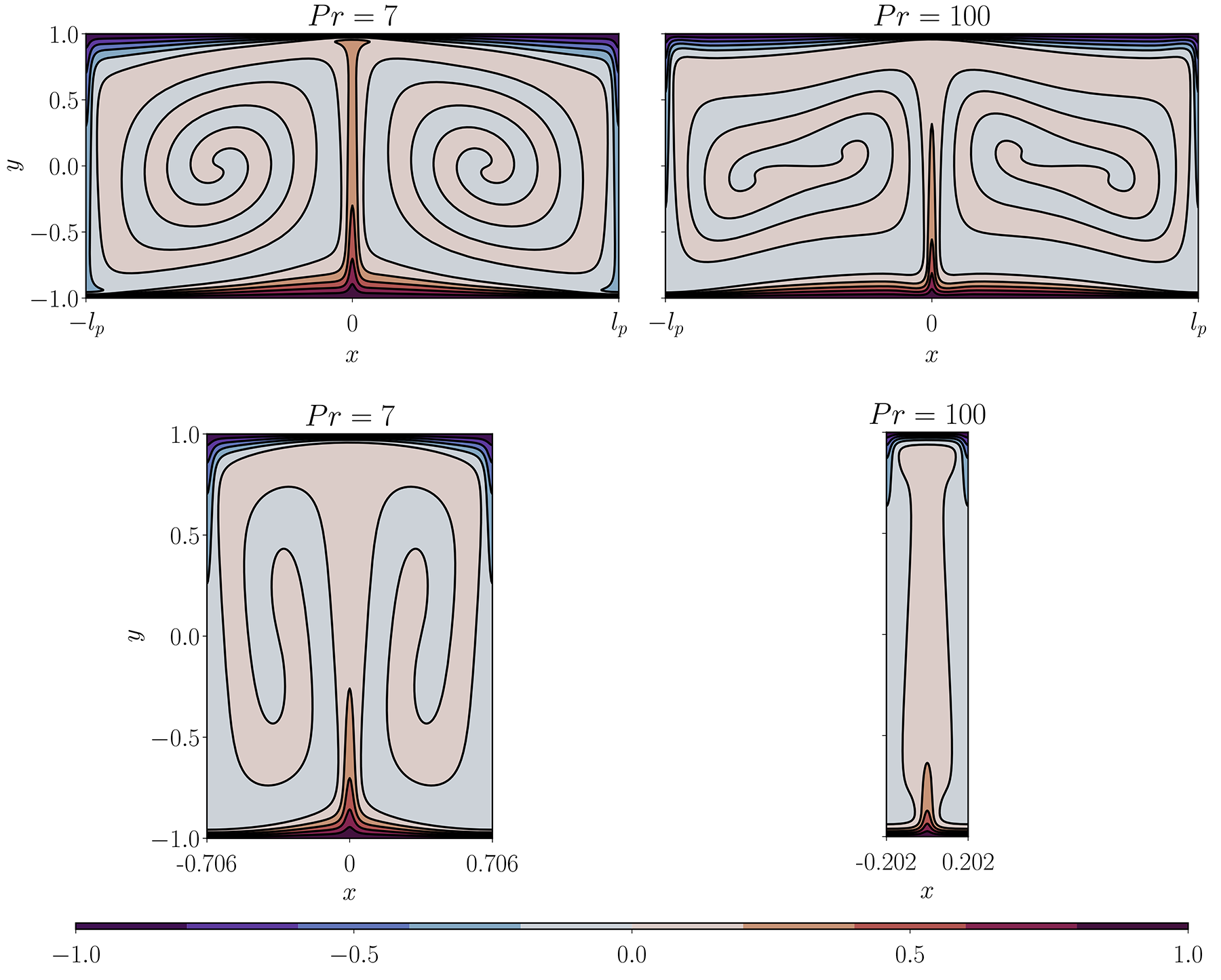}
  \caption{Comparison of primary (top row) and optimal (bottom row) structures at $Pr=7$ and $Pr=100$ at comparable Rayleigh numbers ($Ra=1.05\times 10^{7}$ at $Pr=7$ and $Ra=1.13\times 10^{7}$ at $Pr=100$).}
  \label{fig:prim_opt_comparison}
\end{figure}

\subsection{$Pr = 100$}
\label{subsec:Pr100new}

The results for $Pr = 100$ are similar to those for $Pr =7$ described in the previous section.
On the other hand, for $Pr = 100$, the windowing technique is able to detect embedded optimal solutions for higher $Ra$ up to $Ra \approx 10^8$. The latter result is consistent with Figure~\ref{fig:Nu-Ra-AR10}, showing that the $Pr = 100$ simulation data has values of $Nu$ closer to the optimal values. Furthermore the best-fit scaling $Nu - 1 \propto Ra^\beta$ gives exponent
$\beta \approx 0.293$ for the simulation data, closer to the optimal exponent $\beta \approx 0.311$ than to the primary exponent $\beta \approx 0.227$.
Two possible reasons for the stronger signature of the optimal solution at $Pr = 100$ compared to $Pr =7$ are (i) a larger horizontal scale separation between primary and optimal solutions at $Pr=100$, and (ii) the simpler interior structure of the optimal solution for $Pr=100$. 
Figure~\ref{fig:prim_opt_comparison} shows the structure of the steady solutions at $Ra \approx 10^7$.
The top (bottom) row compares the primary (optimal) solutions for $Pr=7$ and $Pr=100$; the 1st (2nd) column contrasts primary and optimal solutions for fixed $Pr=7$ ($Pr=100$).  One can see that there is a much larger scale separation between primary and optimal solutions for $Pr = 100$ than for $Pr = 7$ at $Ra \approx 10^7$, possibly enhancing our ability to detect the optimal solution at small scales, while observing a signature of the primary solution at larger scales.  
Furthermore,the $Pr=100$ optimal structure has a significantly simpler interior structure than the optimal solution for $Pr=7$. With these pictures in mind, one main objective of this section will be to identify the optimal solution as a persistent small-scale feature for $Pr=100$ in the range $10^5 < Ra \lesssim  10^8$.

\subsubsection{Visualizations and correlation data}
\label{subsubsec:visPr100}

\begin{figure} [h!]
    \centering
    \includegraphics[width=0.75\textwidth]{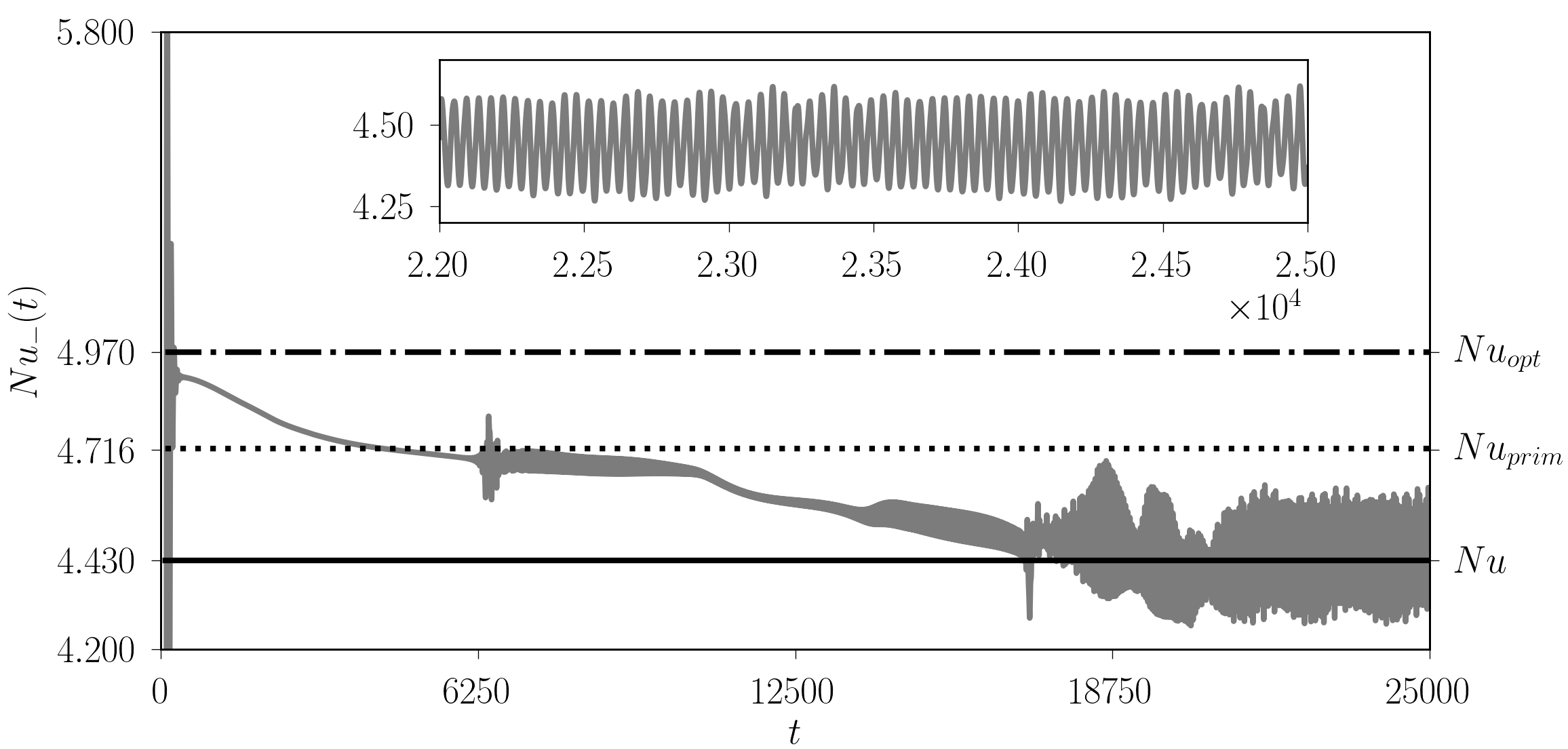}
    \caption{Evolution of $Nu_{-}\lr{t}$ vs.\ $t$ for $Ra = 1.1 \times 10^5$, $Pr=100$. Three time regimes emerge: the flow is quasi-steady ($Q_s$) for $t < 5000$, transient ($\mathcal T$) for $500 < t < 23,000$, and quasi-periodic ($Q_p$) for $t > 23,000$. The inset shows the quasi-periodic stage of the flow for $t > 23,000$.}
    \label{fig:Nu_t_1E5_100}
\end{figure}

Figure \ref{fig:Nu_t_1E5_100} shows the instantaneous Nusselt number $Nu_{-}\lr{t}$ vs. $t$ for $Pr = 100$, $Ra = 1.1 \times 10^5$ ($Re = 8$, $r_c = 64.4)$.
Three distinct regimes are present: (i) a transient regime $t \in [200,5000]$ which is quasi-steady, with decreasing $Nu_{-}\lr{t}$; (ii) a second transient regime $t \in [5000,23000]$; and (iii) a statistically steady state starting at $t \approx 23000$. Regime (iii) was classified as quasi-periodic by observing multiple distinct peaks in the frequency spectrum corresponding to a velocity probe in the middle of the flow field. In the statistically steady regime, the Nusselt number has the value $Nu = 4.43$, compared to the values for the optimal $Nu_{opt} = 4.97$ and the primary solution $Nu_{prim} = 4.716$ (see also Figure~\ref{fig:Nu-Ra-AR10}).

\begin{figure}[h!]
    \centering
    \includegraphics[width=0.9\textwidth]{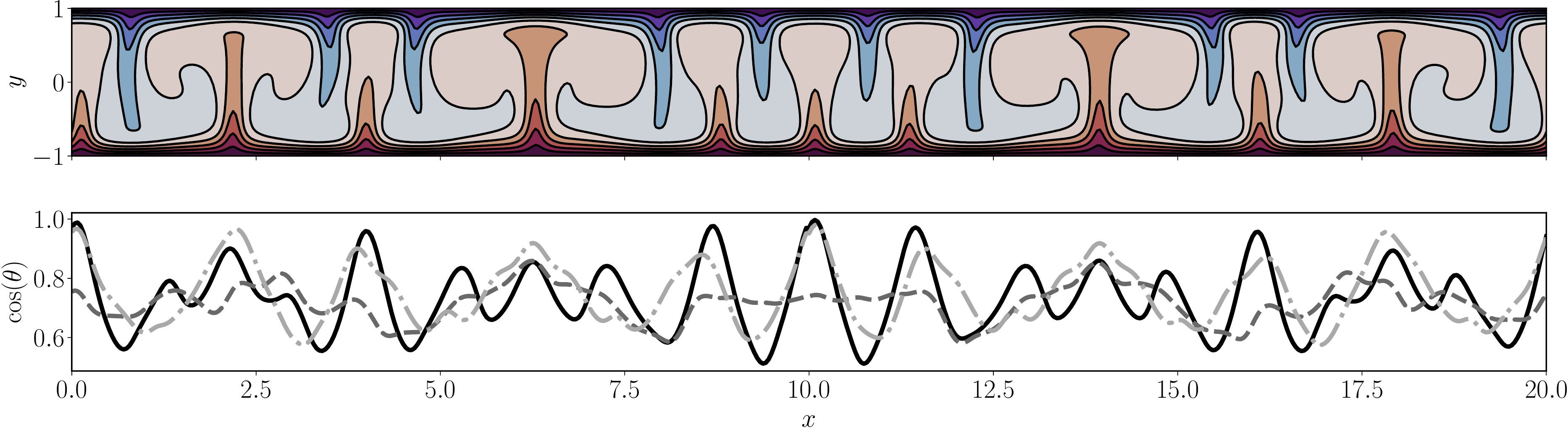}\\[3.0em]
    \includegraphics[width=0.9\linewidth]{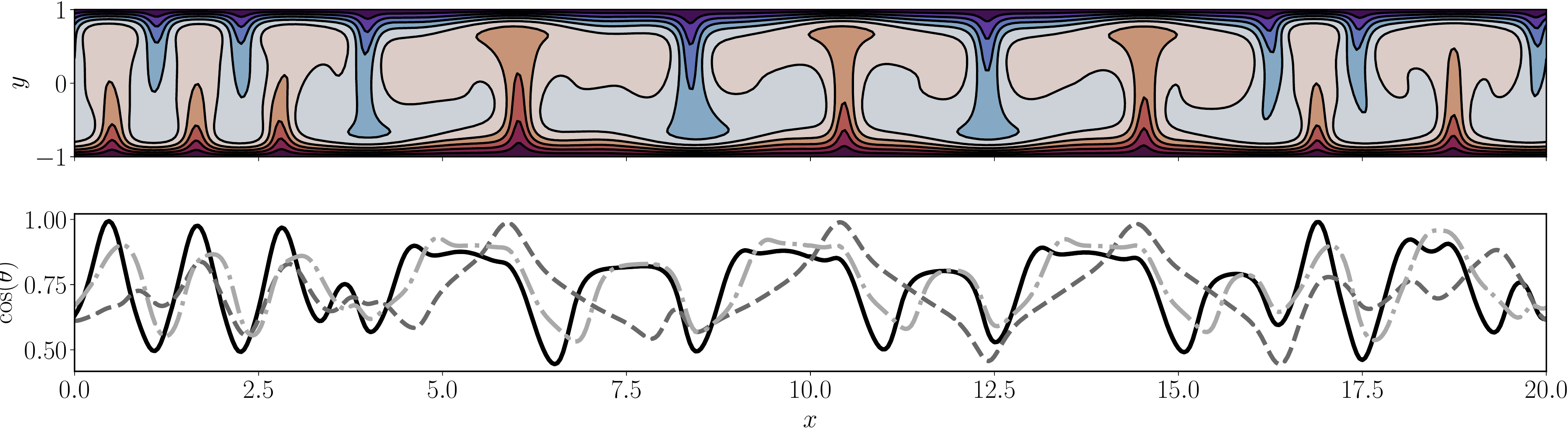}\\[3.0em]
    \includegraphics[width=0.9\linewidth]{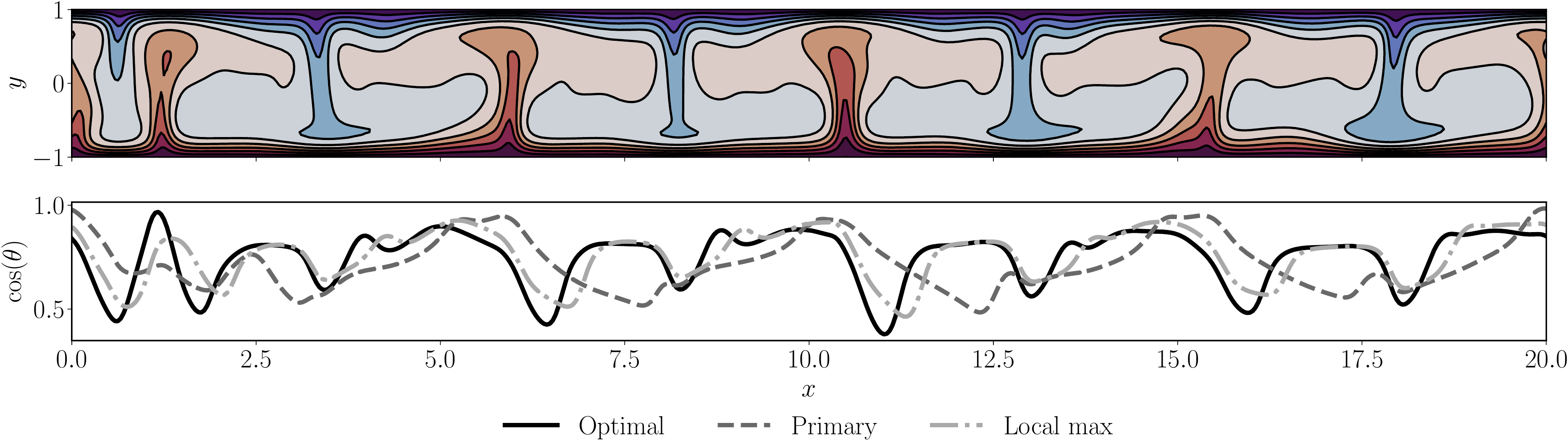}
    \caption{Comparison of temperature fields and correlation coefficients at $Ra= 1.1 \times 10^5$, $Pr= 100$. 
    Top: time $t=1742$ in the quasi-steady regime of the time-developing simulation, with correlation coefficients for the optimal (blue), local maximal (orange) and primary (green) solutions;  Middle: time $t = 11000$ in the transient regime; Bottom: time $t = 24000$ in the quasi-periodic regime.}
    \label{fig:transient-Q_p}
\end{figure}

\begin{figure}[h!]
  \centering
  \includegraphics[width=0.9\textwidth]{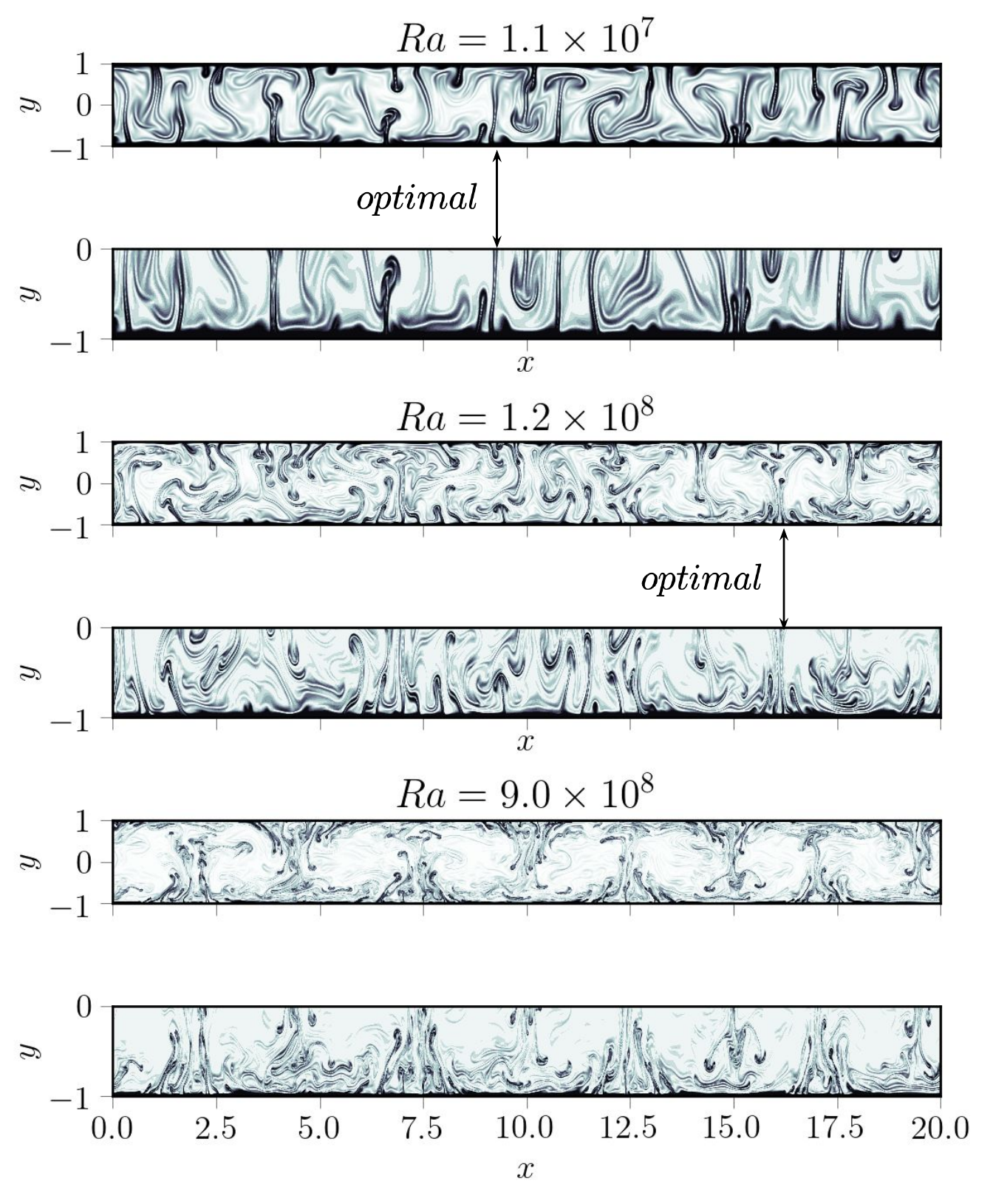}
  \caption{Snapshots of temperature at $Pr = 100$.  For three values of $Ra$, there is a full temperature field (top) and a zoom on the lower half-domain (underneath).  The $Ra=1.1 \times 10^{7}$ and $Ra=1.1\times 10^{8}$ plots are marked with an arrow pointing to a turbulent plume that is well correlated with the optimal solution.  The values of the parameter $\gamma$ in (\ref{eqn:schlieren}) are $\gamma = 1$ in full fields and $\gamma=0.8$ in bottom half-domains.}
  \label{fig:T_100}
\end{figure}

Representative temperature fields for each of the three time regimes are shown in Figure~\ref{fig:transient-Q_p}, along with the value of the correlation $\cos(\theta)$ at each point in the domain for the different solutions: primary, optimal and local-max. Very high correlation coefficients $\cos(\theta) > 0.9$ appear frequently for the flow in the quasi-steady and transient regions.  
Notice, for example, the nearly perfect correlation with the optimal solution at $x=10$, $t = 1742$ (top panel). The optimal solutions have higher incidence than the primary solutions at early times (see Table \ref{tab:stats_prim} and Table~\ref{tab:stats}). In the statistically steady (quasi-periodic) regime, the incidence values for optimal and primary solutions are both roughly $30\%$, but the correlation values are slightly higher for the primary solution, and indeed the dominant structures look like modulated versions of the primary solution (see the bottom panel of Figure~\ref{fig:transient-Q_p}).

As the Rayleigh number is increased beyond transitional values, the Schlieren-type visualizations in Figure~\ref{fig:T_100} show a signature of the primary solution at larger scales, and a boundary layer erupting with many thin plumelets.  
The scale separation is becoming more obvious for higher $Ra$, for example at our highest $Ra \approx 9.0 \times 10^8$.  Our windowing technique and SVD analysis allow us to compare the plumelets to the low-aspect-ratio optimal solutions, comparison of which follows below. For $Ra = 1.1 \times 10^7$ and $Ra = 1.2 \times 10^8$, the arrows in Figure~\ref{fig:T_100} identify two of the plumelets to be analyzed in Section \ref{subsubsec:svd_Pr100}.

\subsubsection{SVD Analysis of Small-Scale Structures}
\label{subsubsec:svd_Pr100}
\begin{figure}[h!]
    \centering
    \includegraphics[width=\textwidth]{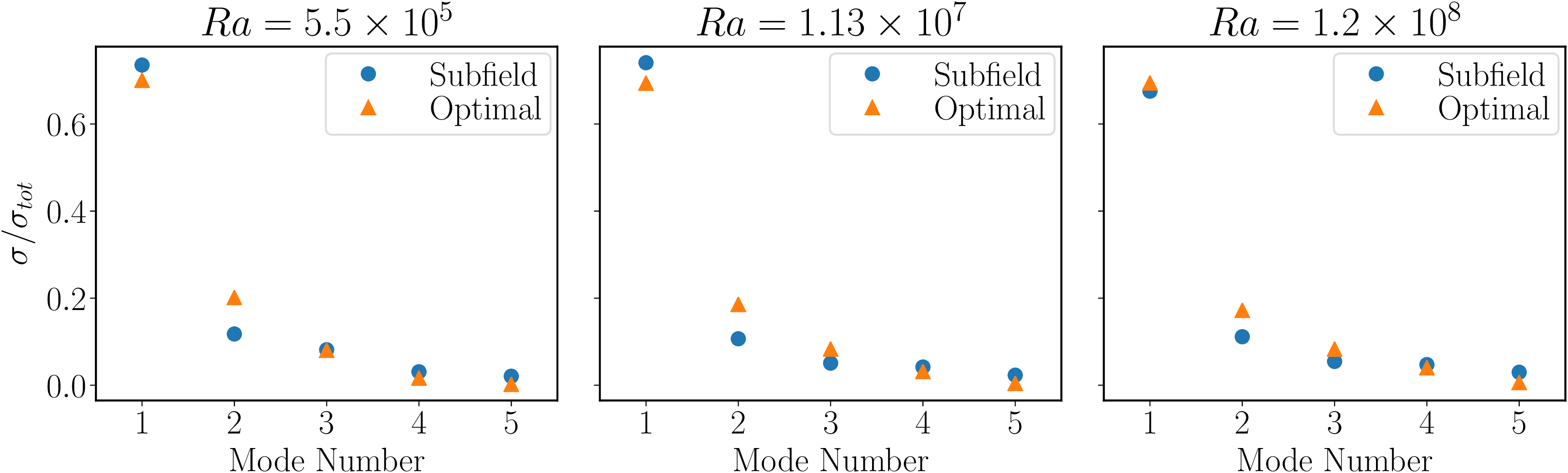}
    \caption{Normalized singular values $\sigma/\sigma_{tot}$, where $\sigma_{tot}$ is the sum of all the singular values. 
    Comparisons are for a simulation sub-box and the optimal solution at (left) $Ra=5.5\times 10^{5}$; (middle) $Ra=1.13\times 10^{7}$; (right) $Ra=1.2\times 10^{8}$.
    The windows corresponding to the simulations are extracted from times in the statistically steady regime.}
    \label{fig:singular_values_Pr100}
\end{figure}

\begin{figure*}[h!]
  \centering
  \includegraphics[width=\textwidth]{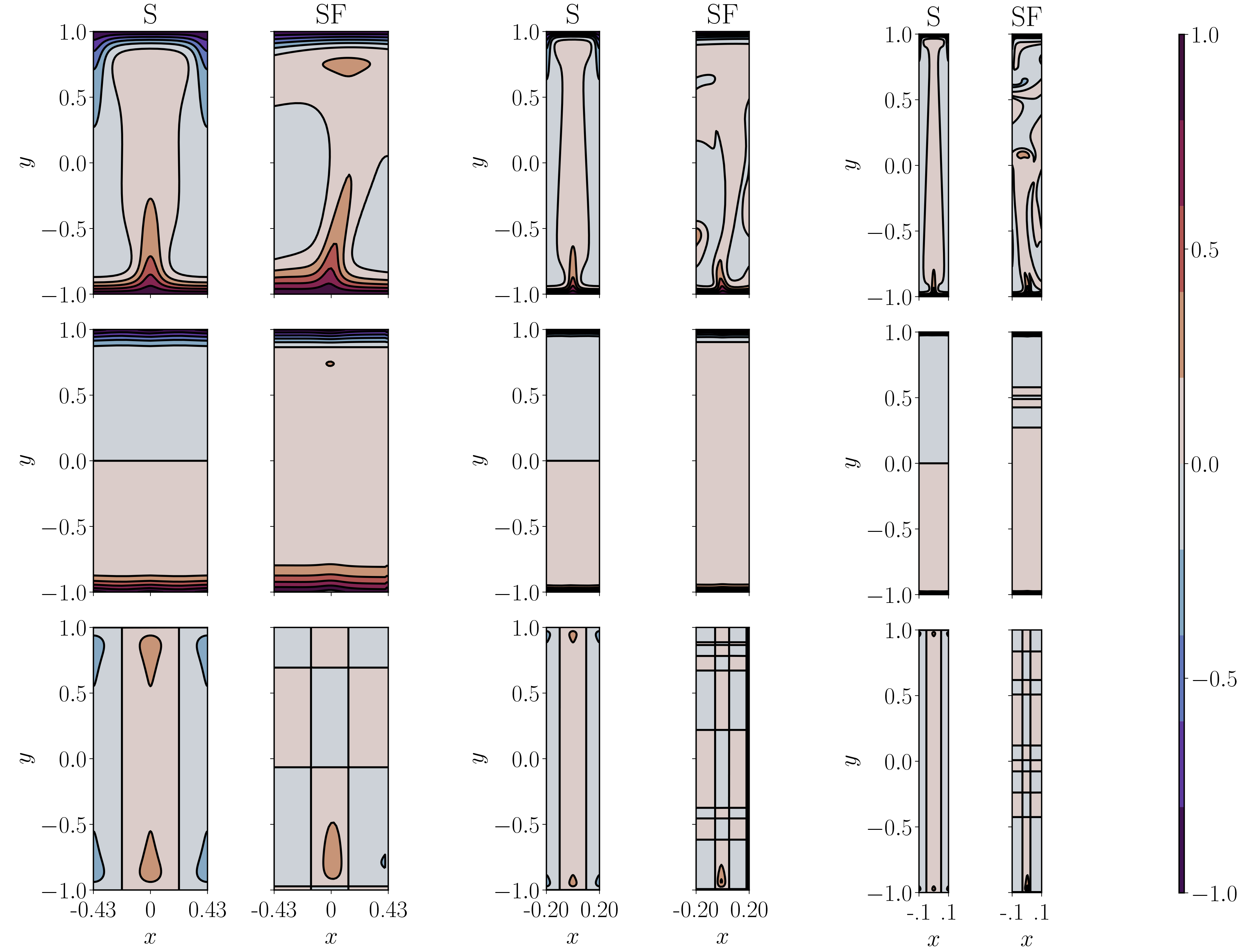}
  \caption{Comparison of optimal coherent structures and structures in the simulation temperature fields at $Ra=5.5 \times 10^5$ (left two plots), $Ra=1.13\times 10^{7}$ (middle two plots), and $Ra=1.2\times 10^{8}$ (right two plots) and $Pr=100$. Pictures of the optimal solutions (S) are true to their aspect ratios and the simulation subfields (SF) have the same aspect ratio.}
  \label{fig:Ra5E5_Pr100}
\end{figure*}

Similarly to section~\ref{sec:svd_Pr7}, in this section we perform the SVD on highly correlated subfields and their corresponding optimal solution. 
Figure~\ref{fig:singular_values_Pr100} compares the singular values of the first 5 SVD modes for three optimal solutions and their corresponding highly-correlated subfields: $Ra=5.5\times 10^{5}$, $Ra=1.13\times 10^{7}$, and $Ra=1.2\times 10^{8}$.
In all three cases, the first 2 modes of the SVD contain greater than $85\%$ of the total energy. Figure~\ref{fig:Ra5E5_Pr100} shows contour plots of the 1st and 2nd SVD modes of the optimal solutions compared with the highly-correlated subfields at $Ra=5.5\times 10^{5}$ (left two columns), $Ra=1.13\times 10^{7}$ (middle two columns), and $Ra=1.2\times 10^{8}$ (right two columns). 
\begin{table*}
  \begin{ruledtabular}
  \centering
  \begin{tabular}{ccccccc} 
  $Ra$  & $\left(\frac{\sigma_1}{\sigma_{tot}}\right)_{s}$ & $\left(\frac{\sigma_1}{\sigma_{tot}}\right)_{sub}$ & Relative Error & $\left(\frac{\sigma_2}{\sigma_{tot}}\right)_{s}$ & $\left(\frac{\sigma_2}{\sigma_{tot}}\right)_{sub}$ & Relative Error \\ [3pt] \hline
  $5.5 \times 10^5$ (optimal) & $0.700$ & $0.735$ & $5.05\%$ & $0.201$ & $0.118$ & $41.5$\% \\
  $1.13 \times 10^7$ (optimal)& $0.694$ & $0.741$ & $6.81\%$ & $0.185$ & $0.106$ & $42.4\%$ \\
  $1.2 \times 10^8$ (optimal) & $0.694$ & $0.675$ & $2.69\%$ & $0.172$ & $0.111$ & $35.1\%$
  \end{tabular}
  \end{ruledtabular}
  \caption{Comparison of relative energy content in SVD modes 1 and 2 for the optimal ($s$) solutions and the sub-fields ($sub$) of the turbulent fields (data and definitions as in Figure \ref{fig:singular_values_Pr100}).}
  \label{tab:stats_svd_Pr100}
\end{table*}
The simulation sub-fields used in the comparison have %$\cos\lr{\theta} = 0.828$
$\cos\lr{\theta} = 0.83$ at $Ra = 5.5\times 10^{5}$, %$\cos\lr{\theta} = 0.797$
$\cos\lr{\theta} = 0.80$ at $Ra = 1.13\times 10^{7}$, and %$\cos\lr{\theta} = 0.790$ 
$\cos\lr{\theta}=0.79$ at $Ra=1.2\times 10^{8}$. Similarly to the $Pr=7$ case, the 1st SVD mode selects the boundary layer. For $Pr=100$, the 1st SVD mode of the optimal solution looks qualitatively similar to the 1st SVD mode of the turbulent sub-field up to $Ra=1.2\times 10^{8}$. Table~\ref{tab:stats_svd_Pr100} presents the relative errors in the 1st and 2nd singular values between the optimal solution and turbulent sub-fields. The relative errors in the first singular values remain remarkably low up to $Ra=1.2\times 10^{8}$. 

For the high $Ra = 1.2 \times 10^8$, it is interesting to note that 
the relative error for SVD mode 1 
is only $2.69\%$ (Table~\ref{tab:stats_svd_Pr100} and Figure~\ref{fig:Ra5E5_Pr100}).
To better appreciate the match, Figure~\ref{fig:distorted} in Appendix~\ref{app:err_plots} is the same comparison between optimal solution and turbulent sub-box presented in the top row, third column of Figure~\ref{fig:Ra5E5_Pr100}, but with a flattened aspect ratio to visualize the small, blunt plumelet emanating from bottom boundary.  The resemblance of the optimal and turbulent structures is quite striking.
Contour plots of the absolute error for the 1st mode are presented in Figure~\ref{fig:Ra100M_Pr100_Opt-Turb_errs} in Appendix~\ref{app:err_plots}. The absolute error in the domain ranges between $10^{-4}$ in the boundary layer to $10^{-1}$ in the bulk region. Typical values are  $3.6\times 10^{-4}$ in the boundary layer [$(x,y) = (0.00488,-0.859)$] and $4\times 10^{-2}$ in the bulk [$(x,y)=(0.00488, -0.00153)$]. We note that at $1.2\times 10^{8}$, the errors in the 2nd SVD mode are considerably larger, reaching order one.

\section{Conclusions}
\label{sec:conclusions}

We investigated a transitional range of Rayleigh numbers $10^5 < Ra \lesssim 10^8$ for $Pr=7$ and $Pr=100$ to find signatures of 2D steady solutions that were first described in~\cite{waleffe2015heat} and~\cite{sondak2015optimal}. These steady solutions satisfy no-slip boundary conditions in the wall-normal direction, periodic boundary conditions in the horizontal direction, and were found by numerical continuation as bifurcations from the conduction state at $Ra \approx 1708$. The primary solution has aspect ratio $\approx 2$, and becomes unstable at $Ra \approx 53,000$ to a time-periodic solution. The optimal solution maximizes heat transport over steady solutions, and has aspect ratio $ < 2$ that decreases with increasing $Ra$. 
A third type of steady solution corresponds to a local maximum of the heat transport, 
also with aspect ratio smaller than $2$.  
These solutions impose mirror symmetry about $x=0$, and generally speaking, their  temperature fields consist of one hot/cold plume pair, with the hot plume emanating from the bottom boundary layer and centered at $x=0$. On the other hand, the three solution types differ with respect to details such as horizontal scale, boundary layer thickness and interior structure.

A domain aspect ratio $\Gamma = 10$ was chosen for the simulations in the current work, allowing multiple copies of 
coherent structures reminiscent of the steady solutions to appear spontaneously from nonlinear interactions, with minimal constraint by the domain dimensions.
The simulation data for $Nu$ vs.\ $Ra$ is almost coincident with the 
data for the primary solution at $Pr=7$ at these relatively low values $10^5 < Ra < 10^8$, especially at the lower end of the range (Figure~\ref{fig:Nu-Ra-AR10}).
For $Pr = 100$, the primary solution becomes less relevant with increasing $Ra$, and the optimal scaling appears to be more dominant. Thus, from the outset, the statistical $Nu$ data strongly suggested a structural relevance of the primary solution for transition to turbulence.
At the same time, the tight upper bound on $Nu$ provided by the optimal solution inspired us ask if the optimal solution might also influence the features of transitional and turbulent data (Figures~\ref{fig:Nu-Ra-AR10}, \ref{fig:Nu_a1E5_7} and \ref{fig:Nu_t_1E5_100}).

Our objective was to establish further links between the simulation data and the steady solutions, beyond the $Nu$ vs.\ $Ra$ information, and for two regimes of $Pr$ represented by $Pr = 7$ and $Pr = 100$. In particular, we used a moving window technique to identify simulation sub-fields with high correction coefficient~\eqref{eq:align}, considering all three types of steady solutions for the comparison at $Ra \approx 10^5$.  As evidenced by Figures~\ref{fig:1E5_100_2},~\ref{fig:1E5_Pr7_t240}, and~\ref{fig:transient-Q_p}, the results are stunning,
%at our lowest $Ra \approx 10^5$, 
with all three steady solutions appearing prominently during transition to statistically steady state. Then, as the flows enter the statistically steady time regime, a modulated version of the primary solution is dominant. For higher $Ra \gtrsim 10^5$, we focused on the primary and optimal solutions, leaving further study of local maximal transport solutions for future work.

For $Ra$ in the two decades $10^7 \leq Ra < 10^9$, Schlieren-type plots -- Figure~\ref{fig:T_7} ($Pr = 7)$ and Figure ~\ref{fig:T_100} ($Pr=100$) -- show a multi-scale horizontal structure in the simulation snapshots.  Visual inspection suggests that the primary solution sets the {\it large} horizontal scale of turbulent plumes, and the intriguing possibility for the optimal solution to impact the {\it small-scale plumelets} that converge together in a primary-scale updraft. We investigated these large-scale and small-scale features using singular value decomposition on highly-correlated sub-fields and, respectively, their corresponding primary and optimal solutions.

Consistent with the $Nu$ vs.\ $Ra$ plots in Figure~\ref{fig:Nu-Ra-AR10}, we found subtle differences in the dominance of the primary solutions at large scales, and the persistence of the optimal solutions at small scales. For $Pr = 7$, the windowing technique and SVD analyses clearly identify the primary solution as embedded within the turbulence at $Ra \approx 10^7$ (Figure~\ref{fig:Ra1E7_Pr7}), while the signature of the optimal solution is much less apparent. We note the lack of scale separation between primary and optimal solutions for $Pr = 7$ at the moderate value $Ra \approx 10^7$ (Figure~\ref{fig:prim_opt_comparison}).  On the other hand, there is a definitive scale separation between primary and optimal solutions for $Pr = 100$, $Ra \approx 10^7$, and the analyses favor the optimal solutions (Figure~\ref{fig:Ra5E5_Pr100}).  
It is conceivable that larger scale separation between primary and optimal solutions will facilitate detection techniques aimed at higher $Ra$ regimes.

\begin{figure}[h!]
  \centering
  \includegraphics[width=0.7\textwidth]{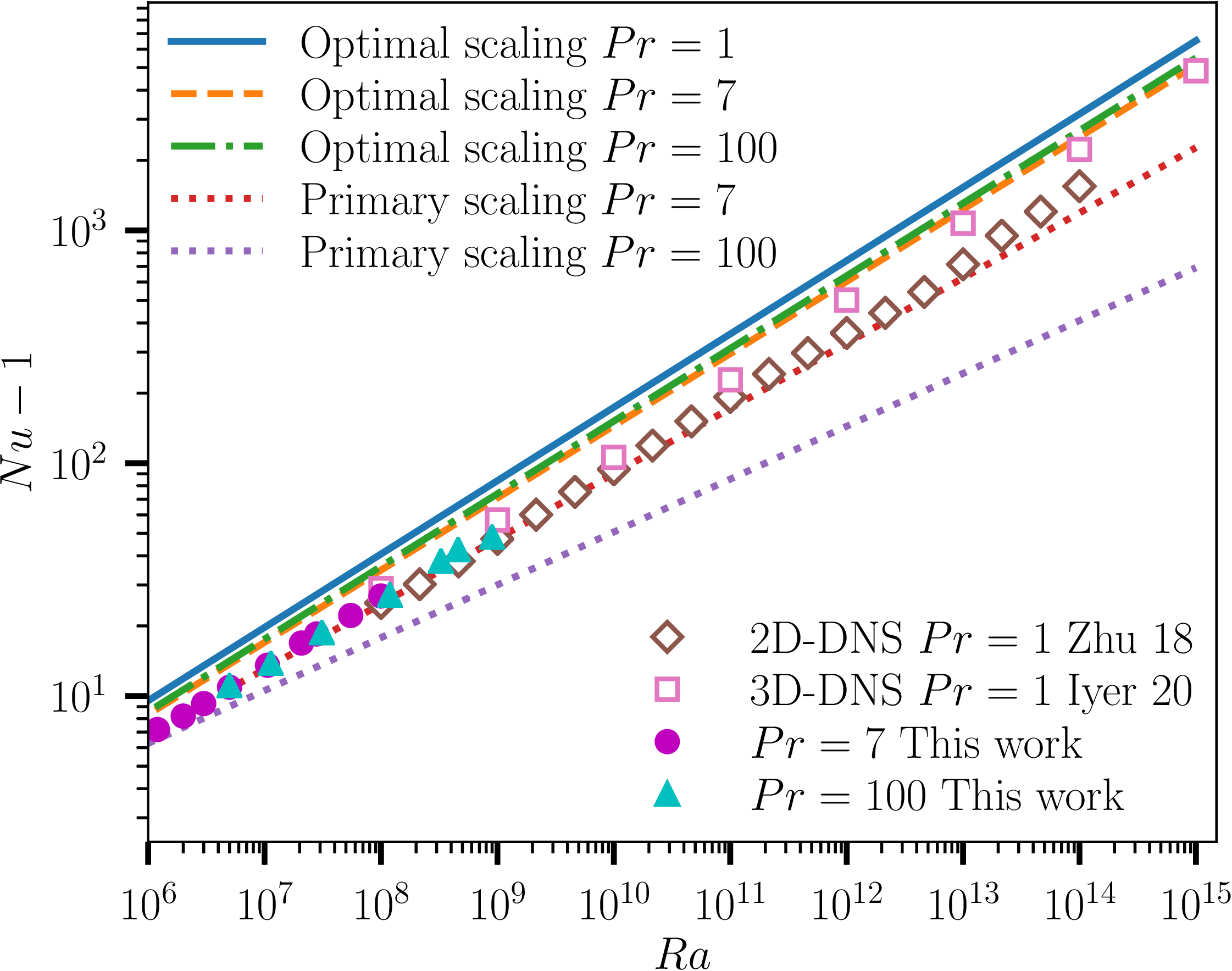}
  \caption{$Nu-1$ vs.\ $Ra$ for various data sets as specified in the legend.}
  \label{fig:speculationzhu}
\end{figure}

An interesting possibility is presented by the high-$Ra$ 2D simulation data of \cite{zhu2018transition} in the range $Ra = [10^8, 10^{14}]$ 
%(see also \cite{johnstondoering2009}\dls{Why?}).  
In a box of aspect ratio $\Gamma = 2$ and for $Pr = 1$, \cite{zhu2018transition} find an approximate scaling $Nu$ vs.\ $Ra^\beta$ with $\beta \approx 0.29$ for $Ra = [10^8,10^{13}]$, beyond which they report a transition to $\beta \approx 0.35$~\cite{zhu2018transition,doering2019absence,zhu2020reply}. Their data is reproduced here in Figure~\ref{fig:speculationzhu}, along with (i) our lower $Ra$ data for $\Gamma = 10$, $Pr = 7$ and $Pr=100$; (ii) the primary solution best-fit scaling at $Pr=7$ extended to the range $Ra = [10^8,10^{15}]$; and (iii) the optimal solution best-fit scaling at $Pr=7$ extended to the range $Ra = [10^8,10^{15}]$. 
One can see a seamless transition between our 2D data for $Ra = [10^5,10^8]$ and the 2D data for $Ra \geq 10^8$ \cite{zhu2018transition}. Furthermore, the continuation of the optimal scaling with $\beta \gtrsim 0.3$ begs the question as to whether or not the transition in 2D exponent
at $Ra = 10^{13}$ \cite{zhu2018transition} could be related to boundary-layer structures with horizontal and vertical scales determined by the optimal solution. 
Thus, 
it would undoubtedly be revealing
to pursue higher values of $Ra$ in 2D, along with application of other types of data analyses (\cite{lee2018model, gonzalez2018deep, fonda2019deep, chen2019machine, wu2020enforcing, pandey2020reservoir}), in the exploration of exact coherent solutions.  

Figure~\ref{fig:speculationzhu} also reproduces the 3D data recently computed for $Pr=1$ up to $Ra = 10^{15}$ in \cite{iyer2020classical}.
One can see that the extended optimal scaling for $Pr=1$ \cite{sondak2015optimal} provides a tight upper bound on the high-$Ra$ 3D data, with %a hint of 
evidence for tendency toward the same scaling exponent as $Ra$ increases.
 %convergence with respect to the exponent of the two scalings.  
As has previously been suggested \cite{waleffe2015heat,sondak2015optimal}, it remains to uncover a possible connection between the 2D optimal solutions and the 3D data.
%which connection seems more likely as their $Nu$ vs.\ $Ra$ relations meet at high $Ra$.
Finally, work to numerically compute exact solutions in 3D is ongoing~\cite{motoki2018maximal, motoki2020multi}, and a longer-term goal is to explore the relevance of 2D and 3D exact solutions to 3D turbulence.

\section*{Acknowledgements}
The authors thank Charles Doering, Andre Souza and Fabian Waleffe for helpful feedback during the development of this work. Fabian Waleffe also contributed suggestions for the written manuscript leading to significant improvements.

\bibliography{RBC}

\appendix

\section{Supplementary Figures}
\label{app:err_plots}

\begin{figure}[h!]
  \centering
  \includegraphics[width=0.8\textwidth]{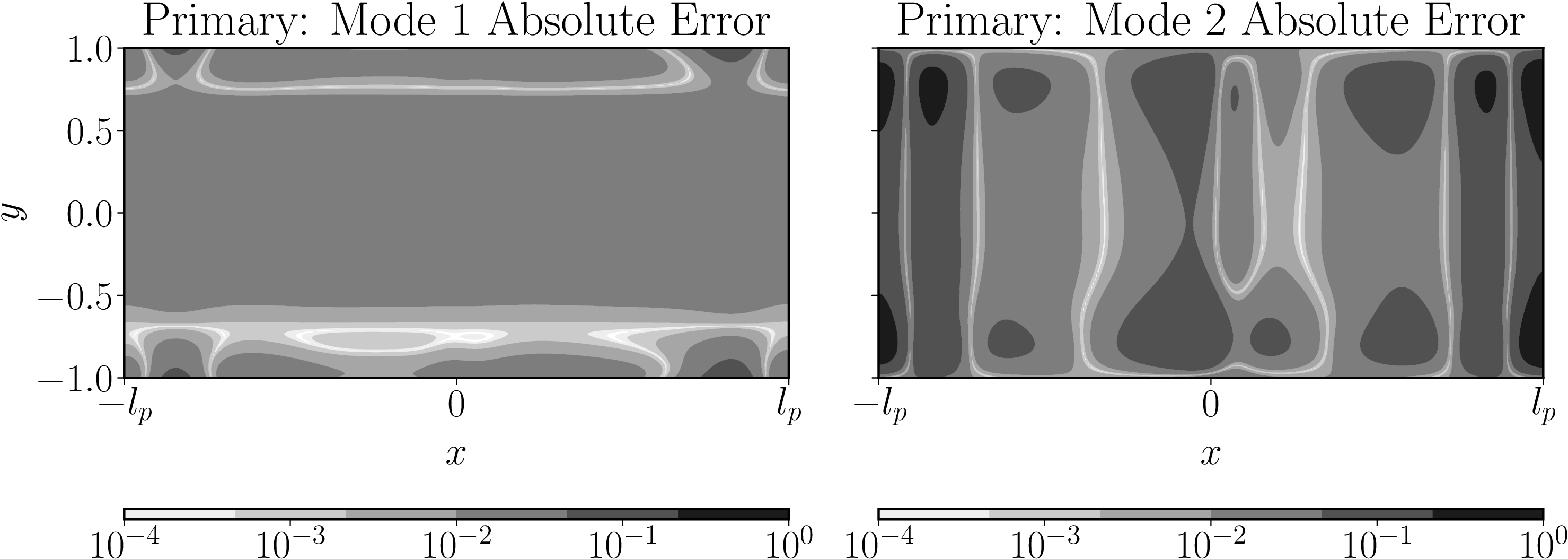}
  \caption{Absolute error of the first (left) and second (right) SVD modes between the primary solution and a turbulent snapshot, both with $Ra = 1.1 \times 10^5$ and $Pr = 7$.}
  \label{fig:Ra100K_Pr7_Prim-Turb_errs}
\end{figure}

\begin{figure}[h!]
  \centering
  \includegraphics[width=0.8\textwidth]{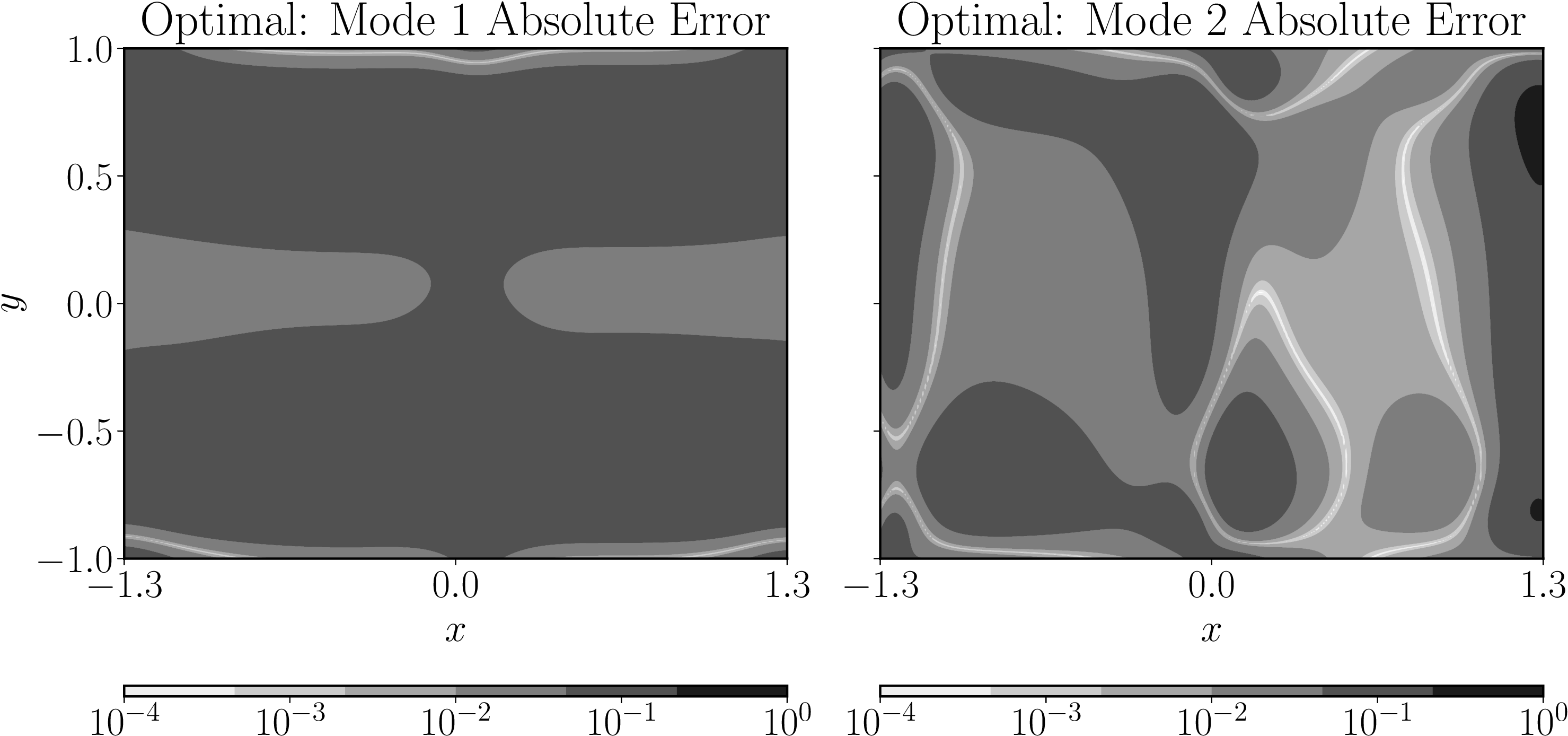}
  \caption{Absolute error of the first (left) and second (right) SVD modes between the optimal solution and a turbulent snapshot, both with $Ra = 1.1 \times 10^5$ and $Pr = 7$.}
  \label{fig:Ra100K_Pr7_Opt-Turb_errs}
\end{figure}

\begin{figure}[h!]
  \centering
  \includegraphics[width=0.8\textwidth]{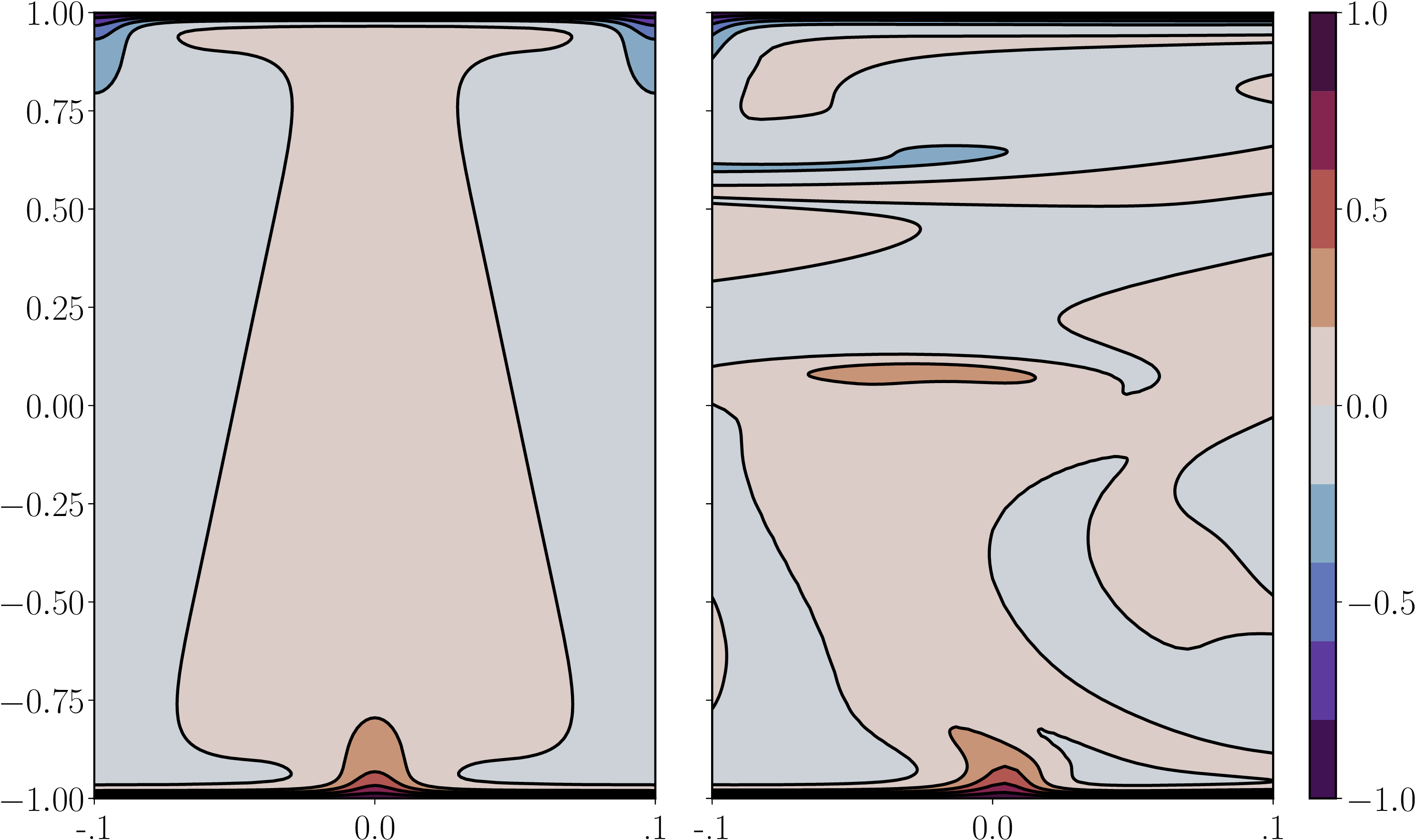}
  \caption{Side-by-side comparison of optimal solution (left) and turbulent temperature sub-field (right) at $Ra=1.2\times 10^{8}$ and $Pr=100$. The aspect ratio has been distorted to highlight the small plumelet at the bottom-center of the window.}
  \label{fig:distorted}
\end{figure}

\begin{figure}[h!]
  \centering
  \includegraphics[width=0.5\textwidth]{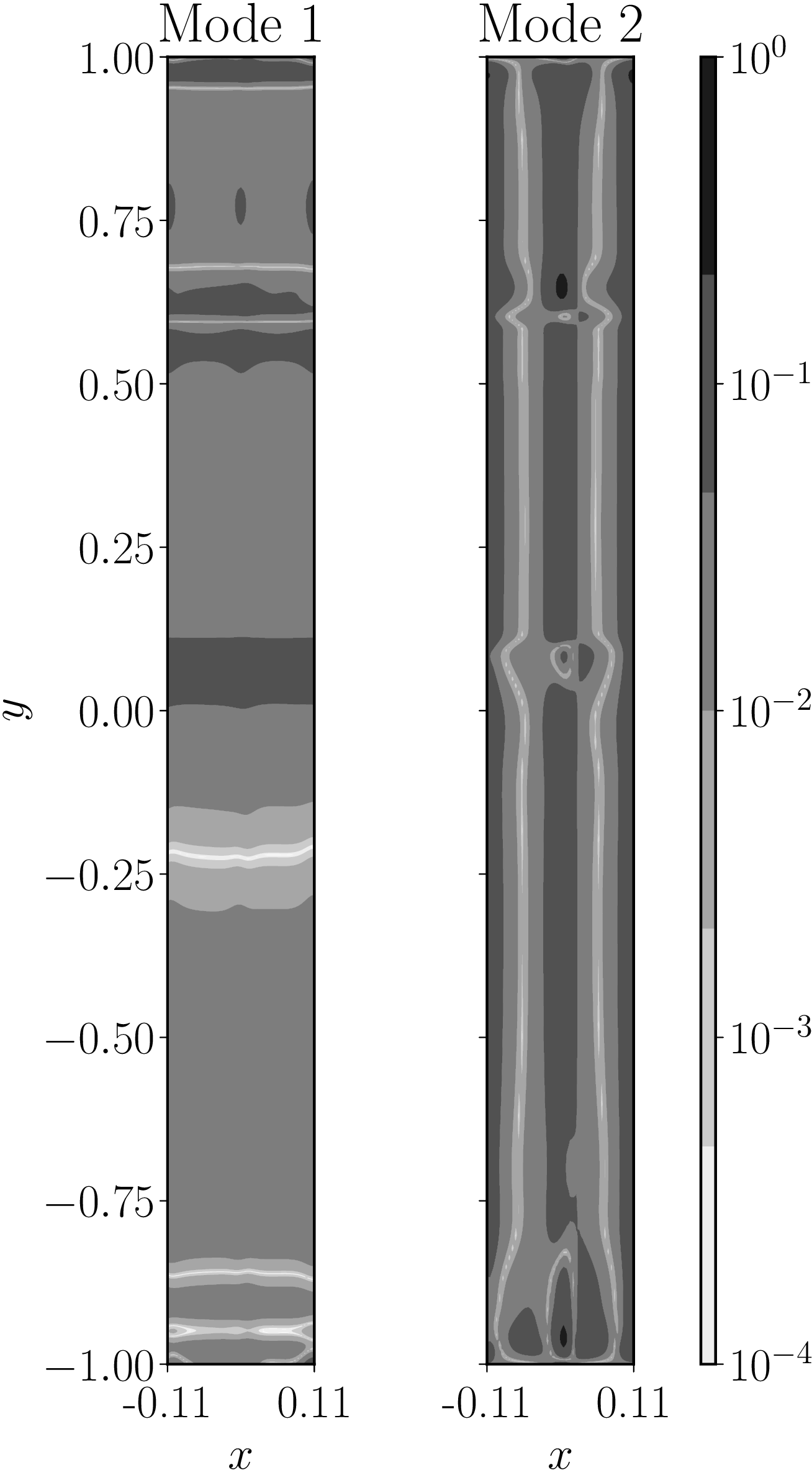}
  \caption{Absolute error of the first (left) and second (right) SVD modes between the optimal solution and a turbulent snapshot at $Ra = 1.2 \times 10^8$ and $Pr = 100$.}
  \label{fig:Ra100M_Pr100_Opt-Turb_errs}
\end{figure}

\end{document}